\documentclass[useAMS,usenatbib, referee]{mn2e}

\def\vaa{\hbox{OGLE004336.91-732637.7}}
\def\vbb{\hbox{OGLE004633.76-731204.3}}
\def\va{\hbox{SMC-SC3}}
\def\vb{\hbox{SMC-SC4}}
\usepackage{graphicx}
 \usepackage{times}

\title{Two Be stars in the Small Magellanic Cloud with B[e]-like properties}

\author[Mennickent et al.]
  {R.E. Mennickent,$^1$\thanks{E-mail: rmennick@astro-udec.cl.
  Based on observations carried out at ESO telescopes: ESO proposal 69.D-0391(A)}
  M. A. Smith$^{2}$, Z. Ko{\l}aczkowski$^{1,3}$\\
  $^1$Universidad de Concepci\'on, Departamento de F\'{\i}sica,
      Casilla 160-C, Concepci\'on, Chile\\
  $^{2}$  Department of Physics, Catholic University of America, Washington, DC 20064, USA; Present
address: Space Telescope Science Institute, 3700\\ San Martin Dr., Baltimore, MD 21218, USA \\  
  $^3$ Instytut Astronomiczny Uniwersytetu Wroclawskiego, Kopernika 11, 51-622 Wroclaw, Poland}
\date{}

\pagerange{\pageref{firstpage}--\pageref{lastpage}} \pubyear{2007}

\def\LaTeX{L\kern-.36em\raise.3ex\hbox{a}\kern-.15em
    T\kern-.1667em\lower.7ex\hbox{E}\kern-.125emX}

\begin{document}

\label{firstpage}

\maketitle

\begin{abstract}

We present an analysis of UVES and MIKE (ground based) high resolution 
and {\it Far Ultraviolet Spectroscopic Explorer} 
spectra of two novel and photometrically variable bright blue stars 
in the SMC; OGLE004336.91-732637.7 (\va) and the periodically occulted
star \vbb (\vb).~ The light curves of these stars exhibit
multiple frequencies, and their spectra are similar. 
The latter are dominated by absorption/emission features produced
in a circumstellar (CS) envelope whereas photospheric features 
are barely visible and forbidden emission lines are not visible.  
Modeling of spectral features indicates similar physical conditions 
for the CS envelope in both stars. An optically thick, slowly
and thermally stratified disk. \vb~ is noteworthy in showing
blue discrete absorption components (``BACs") in their spectra possibly 
indicating the shock interaction between high velocity and low velocity 
material.  Optical spectra from two spectra separated by 5 years 
show little change in the radial velocity over this period. However because
these observations happen to be made in roughly similar phases over their
long period, these suggest only that the stars are not in close binaries. 
We interpret the occultations and additional photometric nearly 
periodic variability in \vb~  as due to covering of the star by a 
density modulation in a sector of a quasi-Keplerian circumstellar disk. 
Altogether, we suggest
that these stars are prototypes of a larger group of stars, which we dub
the ``bgBe's" and discuss their similarities and differences with respect 
to the well known sgB[e] variables. Although \va~ appears to be a 
member the open cluster NGC\,242, the evolutionary context of these
stars is unclear.
\end{abstract}


\section{Introduction}

The evolution of massive stars is an important topic of stellar 
astrophysics, because it provides clues to the mechanisms that feed the 
Galactic medium with the building blocks of future generations of stars. 
During their evolution, massive stars having a solar composition lose some
large fraction of their mass towards the interstellar medium and may 
eventually explode as supernovae near the end of their lifetimes. Owing to 
their weaker winds, the outcomes for metal poor stars are still unsettled
but may be quite different.
Many of the evolutionary stages of massive stars are short-lived and hence 
challenge our ability to characterize them in an evolutionary scheme. 
Detecting objects in these brief phases of evolution should be of great aid 
to test current theories of massive star evolution, especially if they 
represent populations with low metal abundances, a condition that is 
found in the Large and particularly in the Small Magellanic Clouds.
Catalogs of Be star candidates, that is, B stars with extraordinary 
variable light curves in the Magellanic Clouds 
(e.g., Mennickent et al. 2002, hereafter M02 and Sabogal et al. 2006), 
provide excellent material for detailed analysis of possible massive 
pre-main sequence objects and blue stars in rapid evolution stages in 
these galaxies. A census of the pre-main sequence population of massive 
stars in the Clouds has not yet compiled and thus is poorly known. In fact,
only a few objects have recently been placed in this category (Beaulieu 
et al. 1996, Lamers et al. 1999, de Wit et al. 2002, de Wit et al. 2003). 
The general characteristics of these objects are their infrared excesses, 
usual association with nearby nebulosities, their irregular photometric 
variability, and, when well enough studied at high spectroscopic 
resolution,  evidence of infalling circumstellar matter. 
In their study of photometric OGLE light curves of B-type variables 
in the Small Magellanic Cloud, M02 speculated that some of the 
quasi-periodic or periodic variables (Type-3 stars) they discovered might 
be pre-main sequence B stars surrounded by massive gas envelopes.

 Alternatively, we note that other objects described in the M02 and
Sabogal et al. (2006) catalogues have a number of similarities with 
the group of B[e] variables, of which only five are currently known 
in the SMC (Wisziewski et al. 2007; W07).
These are B stars with spectra exhibiting hydrogen line
emissions, $(V-K)$ color excesses between 1.2 and 4 magnitudes,
and occupy regions usually well above the main 
sequence of the H-R Diagram (e.g., Zickgraf et al. 1998, Zickgraf et al. 
1996, and W07).
Lamers et al. (1998) describe these stars as exhibiting moderate to 
large infrared excesses, generally irregular light curves, the presence of
nonspherical circumstellar (CS) structures, winds, and in their spectra
emission lines of permitted Fe\,II and [Fe\,II] and [O\,I]. According to
these authors, the B[e] stars are a highly heterogeneous general population,
consisting of five subclasses. Two of these subclasses, the ``sgB[e]"
(supergiant) stars and the ``unclassified," can exhibit long-term
variations in their light curves.
In particular, ``Algol-like variations" in the light 
curves have been discussed for a few B[e] stars like HD\,45677 and UX\,Ori 
(de Winter \& van den Ancker 1997, Grinin et al. 1994). The variations
of these stars are aperiodic and are generally confined to long intervals 
of a few years.
The sgB[e] stars have a particular set of properties that are largely,
though not completely, similar to our program objects. 
For example, the sgB[e] variations {\it can}
have lower luminosities than supergiant B stars and photometric IR
excesses indicative of the presence of warm CS dust. As for the more general
group of B[e] stars, the sgB[e] stars have spectra that display forbidden 
lines of [Fe\,II] and in some cases molecular lines (Heydari-Malayeri 
1990, Gummersbach et al. 1996, Wisziewski et al. 2007). 
   
    One of the interesting properties of the high luminosity B[e] stars, 
that could be relevant for the objects discussed in this paper, is their
hybrid wind character, that is, a dual wind structure consisting of a
dense slow outflow at the equator and a fast hot one in the polar regions. 
Pelupessy et al. (2000) and others have argued that this structure results 
from a combination of the bistability mechanism for a polar component 
and rotationally induced wind compression for wind efflux emanating from
the stellar equatorial regions.  
Kraus \& Lamers (2003) have presented empirical density 
models for sgB[e] star wind/disk system. Their central argument is that 
in order to permit the presence of appreciable warm dust and CO and TiO
molecules, the equatorial regions must be optically thick and outflowing
slow to effectively shield gas and thereby permit formation of multiatomic 
species. The Kraus \& Lamers models exhibit disk-like structures for which 
loci of constant ionization fan outward from the central plane from distance. 
This permits the detection of multiple types of species by observers at 
intermediate aspects to the star-disk system. Depending on the details 
of the model solution and the ionization level chosen, these contours may or 
may not touch the stellar surface. In particular, the He\,I ionization
surface seems most assured of touching the surface, but because it 
opens to the widest angle it would likely contribute the smallest column
length to CS line formation.
 
   Knowing that any study of M02 variable stars requires spectroscopic 
follow-up to fully understand  their variability, we  selected in year 
2001 two Type-3 stars for a high resolution spectroscopic analysis with 
the ESO Ultraviolet-Visual Echelle Spectrograph (UVES). 
The stars chosen were those in the M02 catalog having visual
magnitudes brighter than 14.2.  There are 8 stars satisfying this 
condition among the 78 SMC Type-3 stars so cataloged.  We selected this 
luminosity simply to assure reasonably high S/N in the spectra 
during the short available observing window we had available with the UVES.
The stars were chosen \vaa~  ($\equiv$ SMC-SC3-63371, MACHO ID 213.15560; 
hereafter \va) and \vbb~ ( $\equiv$ SMC-SC4-67145, MACHO ID 212.15735.6; 
hereafter \vb).  Spectra of both objects showed strong CS gas absorptions.
This fact motivated us to obtain additional optical and also FUV spectroscopy, 
leading to the broad-band study described in this paper.
\va~  is a probable member of the open cluster NGC\,242 
(at a distance of $\sim$ 10" from the cluster center). M02 reports 
for this star two periodicities of 118 and 15 days, whereas \vb~ is an 
emission line object showing occultations (brightness minima)
that recur every 184 days or so, with 
additional nearly periodic photometric variability in time scales of 
$\sim$24 days (Mennickent et al. 2006; hereafter M06).
 
In this paper we examine whether these two stars are pre-main 
sequence B-type, classical Be, or B[e] stars. Of these, the above
sgB[e] possibility will find the closest match.
In Section 2 we give a summary of our observations and present the 
methods of data reduction and analysis. 
In Section 3 we present the methods of spectral line analysis and the 
results of our photometric and spectroscopic analyses in Section 4.  
Discussion and Conclusions are given in Sections 5 and 6. 
The goals of this paper are to elucidate the nature of these two 
rather luminous B-type variables, to discuss their unique array of CS 
lines, and to pave the way to assign an evolutionary state of a 
potentially new class of Be stars related to the sgB[e] stars.

\section{Observations and data reduction}

High resolution spectra at resolving power $\sim$ 40\,000 were obtained
on 2001 May 17 with the UVES spectrograph in dichroic modes at the UT\,2 
telescope in the ESO Paranal Observatory, Chile. The four CCD chips allowed
sampling the spectral ranges of  3050-3870 \AA, 4780-5756 \AA, 5832-6807 \AA~  
and 6704-8517 \AA.  A slit width of 1" was used, allowing us to
obtain good spectra in 600 sec exposures. Wavelength calibrations were 
performed with the UVES pipeline at the telescope. The $rms$ of the 
wavelength calibration function regarding Th-He-Ar comparison lines 
was a $\sim$0.01 km\,s$^{-1}$.
The spectra were normalized to the continuum and no flux calibration 
neither telluric correction  were 
intended. Additional spectra were obtained with the {\it Magellan 
Inamori Kyocera Echelle} (MIKE) spectrograph at the Clay telescope in 
Las Campanas Observatory, Chile in 2007 November 09. 
For this double echelle spectrograph the wavelength
range was 3390$-$4965 \AA~(blue part) and 4974$-$9407 \AA~ (red part) 
whereas the slit width was 0.7". The resolving power and dispersion 
were about 50\,000 and  0.02 \AA\,pixel$^{-1}$ (for the blue part) and 
40\,000 and  0.05 \AA\,pixel$^{-1}$ (for the red part). 
The $rms$ of the calibration function for these spectra was a 
$\sim$0.01 km\,s$^{-1}$.  The spectra were reduced and calibrated with IRAF. 
A summary of the observations is given in Table\,1.

In addition,  far-UV spectra were obtained with the {\it Far Ultraviolet
Spectroscopic Explorer} ({\it FUSE}), as detailed in Table\,2. These data cover
the continuous wavelength of approximately $\lambda \lambda$929-1187 \AA~ by
means of eight independent MAMA detector segments. The spectral resolving
power among these varies but is typically 15\,000. The spectra were reduced
with the CalFUSE version 3.1.8 pipeline system, which was very similar to
the last version used for the final reprocessing of the {\it FUSE} archive.

\section{Methodology}

  Much of the analysis in this paper relies on spectral line 
synthesis, so we first describe the methods used for this analysis.
We have assumed log\,g = 3.5, appropriate for luminosity class I, early 
B-type stars (Crowther et al. 2006),  and a metallicity of 0.2$\times$ solar 
for our objects (e.g., Dolphin et al. 2001, Mighell et al. 1998, Dufton et 
al.  2005).  The spectra were analyzed using Hubeny/Lanz {\sc synspec} 
and {\sc circus} programs (Hubeny, Lanz, \& Jeffery 1994, Hubeny \& Heap 
1996). For {\sc circus} analysis an input temperature for an intervening
circumstellar structure, $T_{c}$, and other parameters must be assumed 
as well.  The Fe\,II and Si\,II lines below 
suggest a microturbulence $\xi$ in the range of 5-10 km\,s$^{-1},$
and we assume the disk is comoving at a constant velocity. 
We also posited that the foreground ``cloud" (CS wind/disk) covers
the star completely in our models.  
These assumptions require the matching of the computed equivalent widths 
to the observations using a two-parameter fit in $T_{c}$ and column density. 
It was readily apparent, both from the columns needed to fit the
observed line strengths as well as from strength ratios of lines 
arising from the same ion, that the CS lines are optically thick.  
In some cases the lines are observed to have peak equivalent widths 
(EWs) as large as we can compute them even from assumed optimal conditions. 
This suggests that they are indeed formed in conditions near the optimal 
$T_{c}$ and  column density values. However, this assumption does impose 
errors, generally as underestimates in the column density. Thus, if the 
temperature other than the optimal value, a larger column density will 
be required in the line fitting. This dependence emphasizes correlated 
errors in the determination of the column density and assumed $T_{c}$. 
Most of our analysis is for absorption features.  However, for the 
cases of far-UV C\,III multiplet, the optical He\,I, and strong 
Fe\,II lines, we fit line profiles with a medium radiating in 
emission, that is, as if were formed in lines of sight other than
than those intersecting the star disk.

\begin{table*}
\centering
 \caption{Summary of UVES/MIKE observations. The MJD numbers at mid exposure are given.
 The ephemeris for the occultations of  \vb~  is 2\,450\,709.9(2) + 183.9(1.0)\,E. For \va~ we use
 2\,450\,000.0 + 119\,E. }
 \begin{tabular}{@{}ccccccccc@{}}  
object & instrument &UT-date   &    airmass & ccd &   grating & exptime (s) &mjd-obs &phase \\
\hline
\va~ &UVES &2001/05/17&  1.96& CCID-20 &CD\#3&  600 & 52411.39285 &0.26\\
\va~ &UVES &2001/05/17&  1.96& CCD-44  &CD\#1&  600 & 52411.39287 &0.26\\
\va~ &UVES &2001/05/17&  1.91& CCID-20 &CD\#3&  600 & 52411.40188 &0.26\\
\va~ &UVES &2001/05/17&  1.91& CCD-44  &CD\#1&  600 & 52411.40189 &0.26\\
\vb~ &UVES &2001/05/17&  1.86& CCID-20 &CD\#3&  600 & 52411.41265 &0.26\\
\vb~ &UVES &2001/05/17&  1.86& CCD-44  &CD\#1&  600 & 52411.41268&0.26\\
\vb~ &UVES &2001/05/17&  1.81& CCID-20 &CD\#4&  600 & 52411.42434 &0.26\\
\vb~ &UVES &2001/05/17&  1.81& CCD-44  &CD\#1&  600 & 52411.42435&0.26\\
\va~ &MIKE &2007/11/09&  1.43& Lincoln-Labs+SITE ST-002A &echelle&  500 &54413.03654  &0.08            \\
\vb~ &MIKE &2007/11/09&  1.42& Lincoln-Labs+SITE ST-002A&echelle&  500 &54413.04421   &0.14            \\
 \hline
\end{tabular}
\end{table*}

\begin{table}
\centering
 \caption{Summary of {\it FUSE} Observations. Phases refers to the ephemeris given in Table\,1.
}
 \begin{tabular}{@{}cccccc@{}}  
object & UT-date  &UT-start & exptime (s) &mjd-obs &phase\\
\hline
\va~   &  2006/10/05&04:05:06& 2430  &54013.18427083 &0.72\\
\vb~   &  2006/11/30&11:36:26& 2736  &54069.49946759 &0.27\\
 \hline
\end{tabular}
\end{table}

\section{Results}

\subsection{Photometry}

  Magnitudes and colors for both program stars are given in Table 3. Their
derived colors are consistent with reddened early B or late O type stars, 
but the reddening cannot be estimated well because most of it occurs
in the stars CS disks. For example, in Table\,3
a crude estimate based on the $Q$ reddening-free parameter gives
a color excess E($B-V$)= 0.27 and 0.29 magnitudes for \va~ and \vb, based
on their assumed intrinsic $B-V$ colors, ``BV0."~
An alternate estimate of the influence of reddening can be obtained by
using the Exposure Time Calculator (ETC), a tool used by proposers to
the Hubble Space Telescope Guest Observer program.
It gives consistent count rates for the measured {\it FUSE} 
flux at 1180\,\AA~ and the $V$ magnitudes, leading to {\it equivalent} 
E($B-V$) values of 0.38 and 0.48 magnitudes, respectively. 
However, because the Hubble ETC tool likewise assumes a normal 
Galactic reddening curve from the
far-UV to optical regimes, whereas the far-UV dimming in our case is most
likely due to a complicated array of atomic and molecular line absorptions, 
we can claim the latter only as estimates useful for comparing fluxes in
the visual and far-UV wavelength regimes.

  For the purposes of comparing our program stars to B[e] stars, it is
useful to know that the observed ($V-K$) colors for our program 
stars are about +1.0 magnitude (see Table\,3).
Anticipating results below for the intrinsic ($V-K$) color of a B0\,I 
star, -0.7 (Tokunaga 1999), the IR reddening for our two program objects 
is about 1.7 magnitudes. Comparing this with the range of 1.2-4 
magnitudes for this quantity noted above for the 5 known sgB[e] stars 
in the SMC, we see that the program stars fall on the low side but 
still within this range.

In Fig.\,1 we show the color-magnitude diagram for \va~ and for 
neighboring stars present in their field. We investigated the light 
curves for the stars labelled 1, 2 and 3 and they are quite constant. 
The figure also depicts isochrones for the cluster NGC\,242 
considering the parameters indicated in the figure caption.  
It is clear that especially those bright objects lie close to the
$\log\,t$ =  7.8 isochrone. 
The Fig.\,1 also depicts the observed position of \va.~ Its true 
location the HR Diagram should be projected back along the indicated 
reddening vector.

Deep exposures of sky images surrounding the two program objects 
reveal no suggestion of nebulosity. However,
during our investigation we discovered that \va~ has a visual
companion (the small bump at the NW of \va~ in Fig.\,1) about 1" from
\va.~ We analyzed its OGLE light curve as part of this study and 
discovered that this visual companion is 
an eclipsing binary with an orbital period of 2.96934 days, $V$= 16.776, 
$V-I$= -0.083 and range of variability in $I$ equal to  0.77 mag. 
This nearby star is not in the catalog of eclipsing binaries  in the $SMC$ 
(Bayne et al. 2002).  It is unlikely that this star influences the 
periodicities found in the OGLE light curve of \va~. 

The O-C diagram for \va~ using OGLE\,II and OGLE\,III data shows that 
the 15-day cycle is not constant (Fig.\,2a).  
Adding the MACHO light curves (Allsman \& Axelrod 2001), we found that 
the amplitude of the 15-day periodicity is larger in the blue band
(Fig.\,2b). On the other hand, the O-C diagram for \vb~ (Fig.\,3a) suggests 
that this star undergoes quasi-periodic occultations. In fact, the $\sim$184 
day minima are irregular (Fig.\,3b), of larger amplitude at the blue 
band and the O-C diagram indicates that the period varies. 
All these features, along with the presence of additional 
quasiperiodic variability in time scales of 24 days (M06), prompt 
a suspicion that the eclipses are due to obscuration of the star 
by a rotating inhomogeneity in the CS cloud. The inference that at 
least a dense CS cloud exists around this star is borne out by 
our spectroscopic analysis below.

\begin{figure*}
\scalebox{1}[1]{\includegraphics[angle=0,width=15cm]{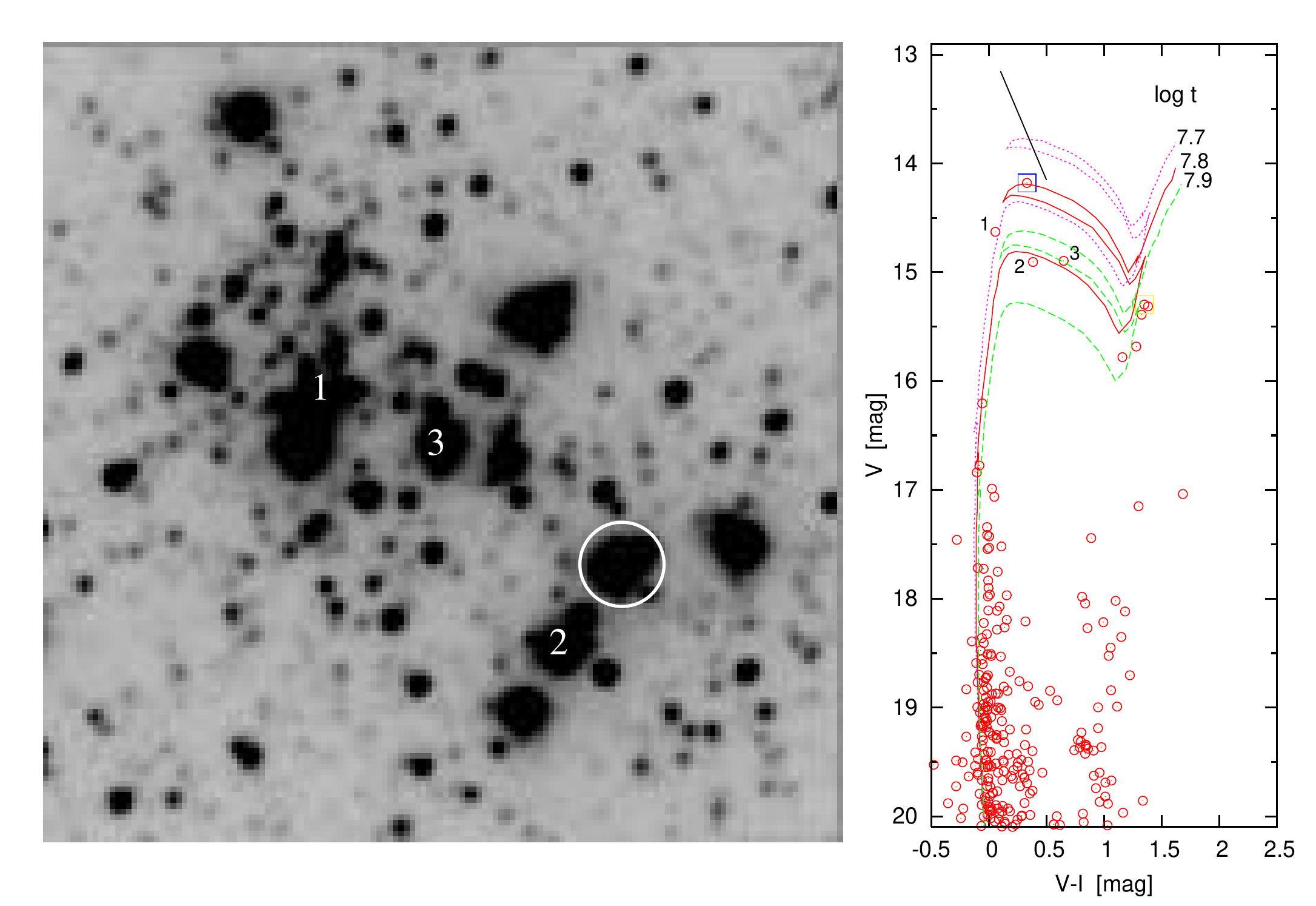}}
\caption{The left picture is a  finding chart of \va~ centered on the 
cluster NGC\,242 (1' $\times$ 1' subframe of the reference $I$-band 
image from the OGLE-II survey).   North is up and East left. \va~  is 
indicated by a circle and three reference stars are labeled.  Right panel 
shows the CMD for the same region. The CMD is based on the OGLE\,II standard 
$VI$ photometry in the SMC (Udalski et al. 1998). Three isochrones are also 
presented (Bertelli et al. 1994) for stars with metallicity $Z$= 0.004
and ages ($\log t$) labeled in the picture. 
We adopted these isochrones using $V-M_{v}$= 19.0, $E(V-I)$= 0.08 and
$A_{v}= 2.5*E(V-I)$. \va~  is marked with an open square and the reference 
stars with circles.  The reddening vector is indicated by the upper 
oblique line.}
  \label{1}
\end{figure*}

\begin{figure}
\scalebox{1}[1]{\includegraphics[angle=0,width=8cm]{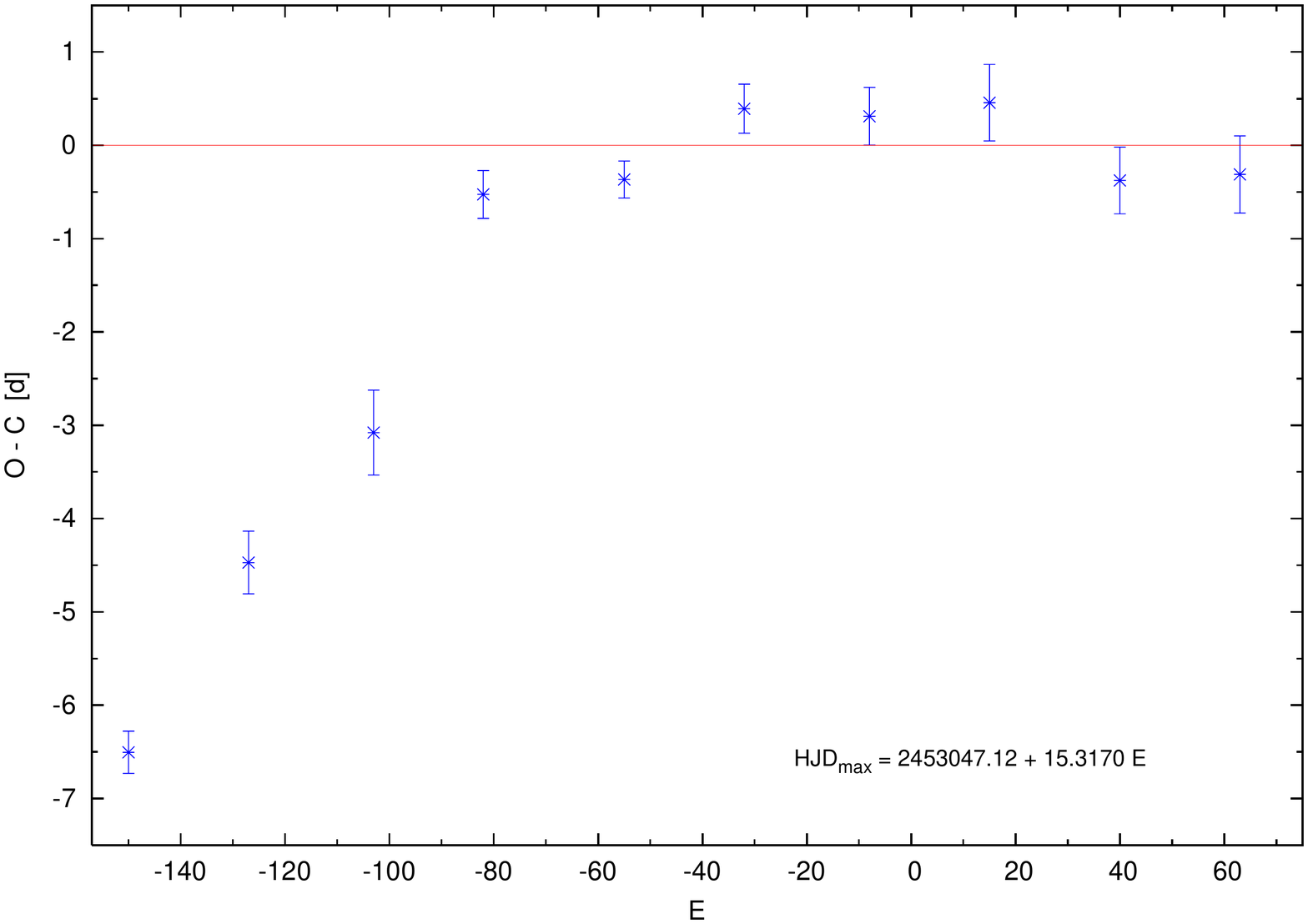}}
\scalebox{1}[1]{\includegraphics[angle=0,width=8cm]{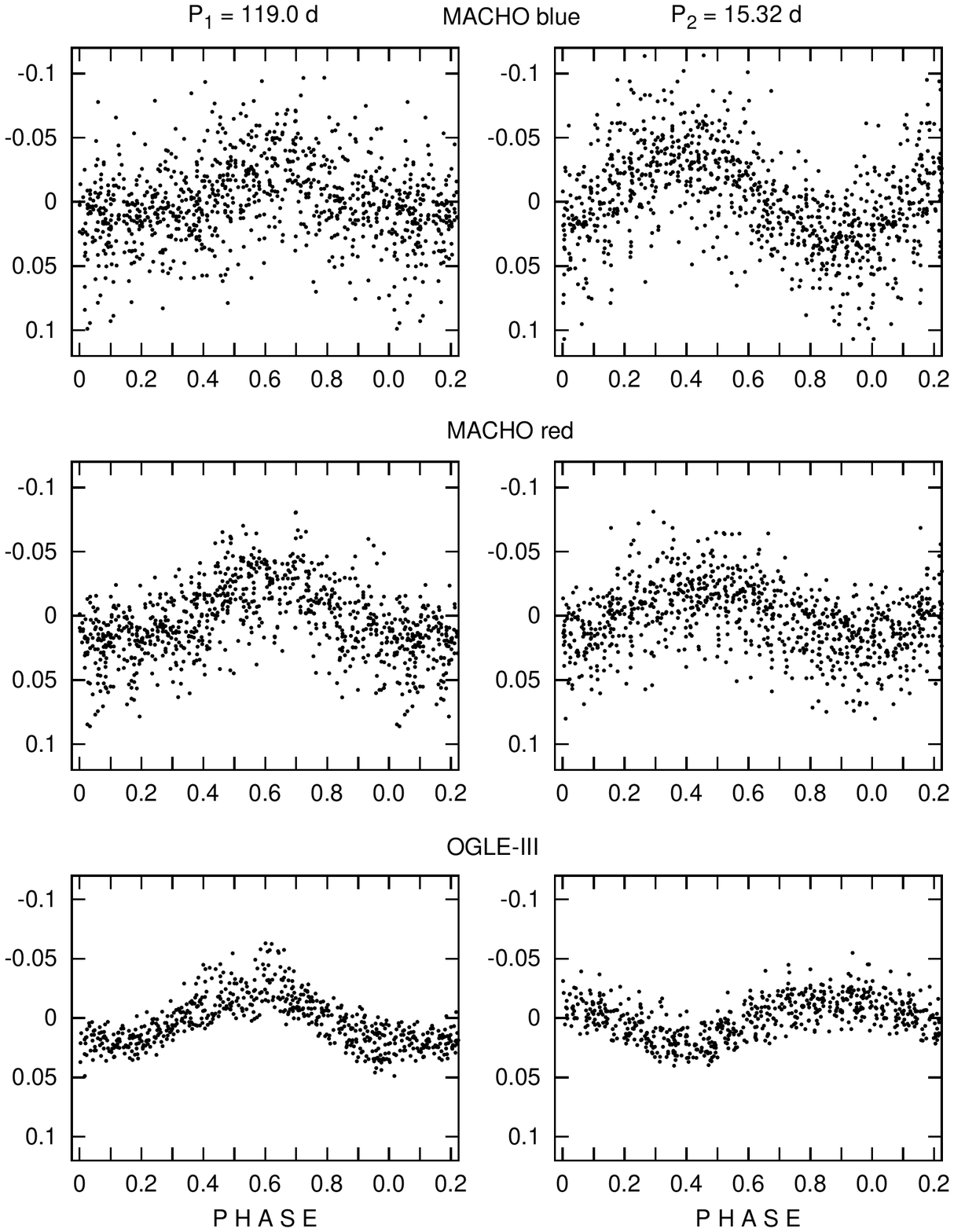}}
\caption{Left: O-C diagram for the 15-d periodicity of \va. Each point was calculated for one season (number of observations 
per season is from 60 to 100) and the model fit includes both periods.  The primary period $P_{1}$= 119 d has non-sinusoidal shape, therefore 
its first harmonic was added (2 $\times f_{1}$) in the final fit. 
The ephemeris was calculated using the second period ($P_{2}$= 15.3170 d) 
from the OGLE\,III data, since for these epochs it seems to be constant. 
Right: MACHO and OGLE\,III light-curves of both periods detected in 
\va~ are folded with $HJD_{0}$ = 2\,450\,000.0. 
This figure illustrates the $P_{2}$ period change and the amplitude 
difference between bands. Note that $P_{1}$ has a non-sinusoidal shape. }
  \label{2}
\end{figure}

\begin{figure}
\scalebox{1}[1]{\includegraphics[angle=0,width=8cm]{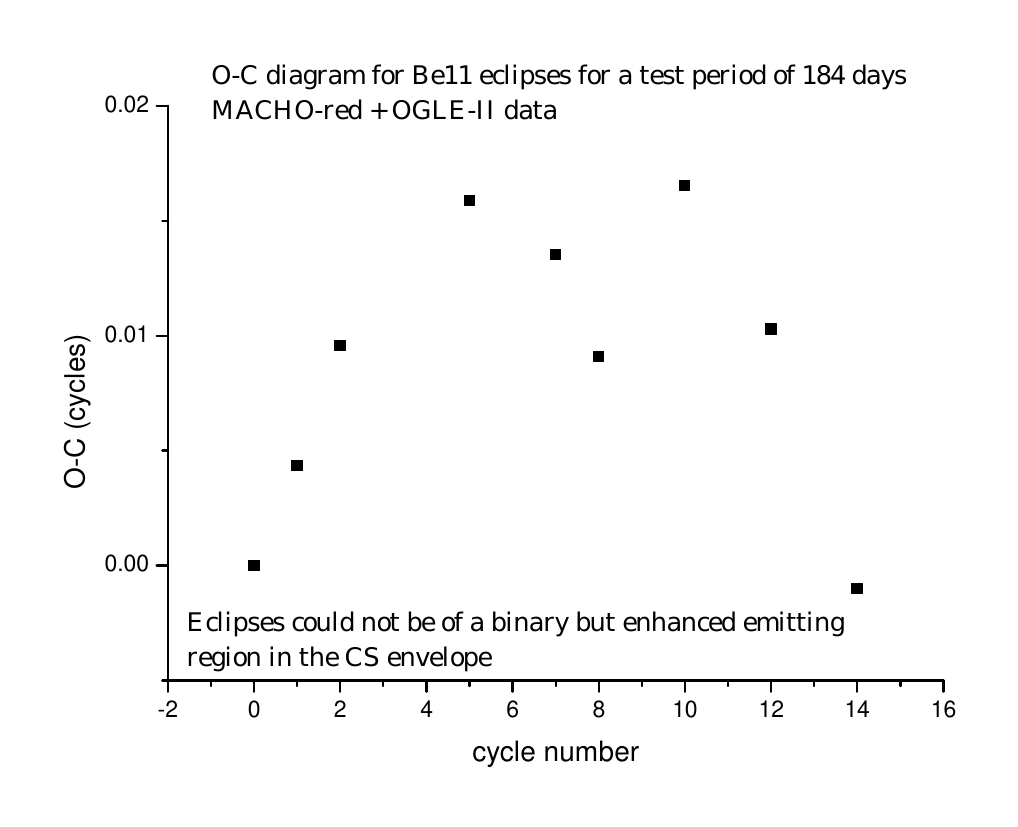}}
\scalebox{1}[1]{\includegraphics[angle=0,width=8cm]{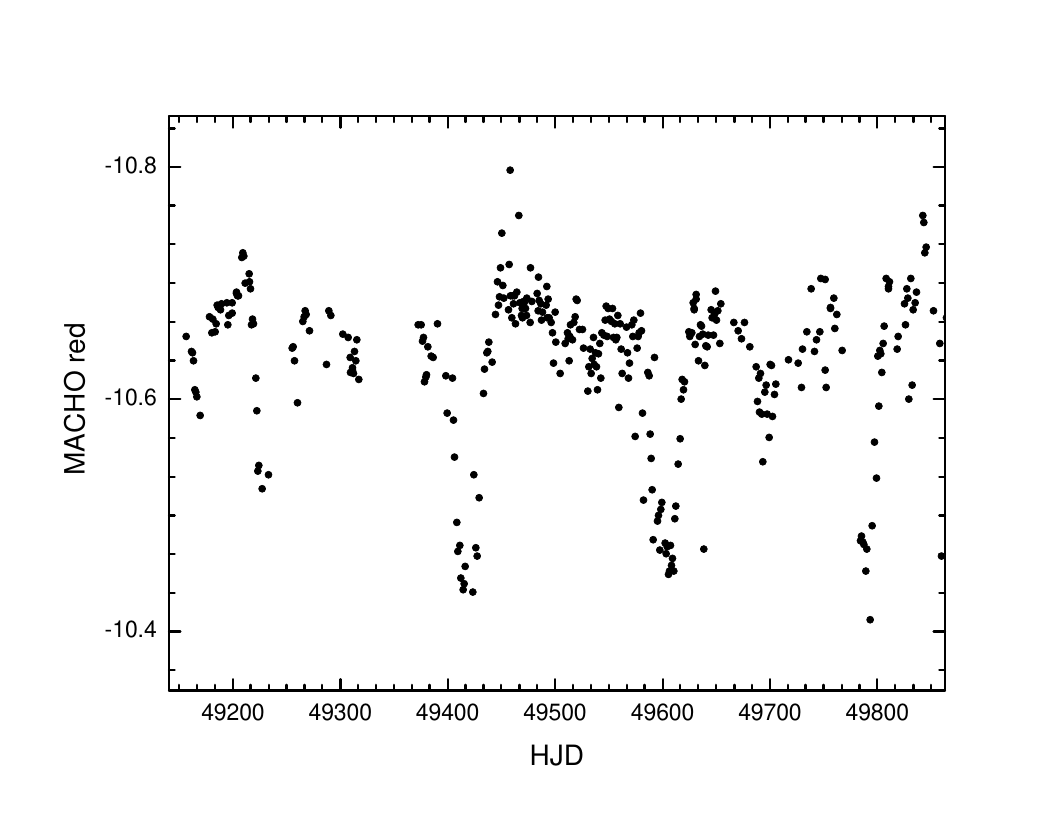}}
\caption{Left: O-C diagram for \vb. Right: Variability of eclipses in \vb.}
  \label{3}
\end{figure}

\begin{table*}
\centering
 \caption{$UBVI$ magnitudes from Zaritsky et al. (2002) and OGLE photometry 
are given, along with reddening-free $Q$ (based on OGLE and Zaristky U 
magnitude), Q2 (entirely based on Zaritsky photometry) 
 and $B-V$ indices (based on the Q factor) and the color excess according 
to the Johnson \& Morgan (1953) calibration.}
 \begin{tabular}{@{}cccccccccccc@{}}
  \hline 
Object &U &B& V& I& V(OGLE)& BV(OGLE)& VI(OGLE) &Q & Q2& BV0 & EBV \\
\hline
\va~ & 14.262(40)& 14.392(33)& 14.085(26)& 13.729(23) & 14.18&  0.181& 0.331& -0.23& -0.35& -0.09& 0.27\\
\vb~ & 14.179(89)& 14.205(21)& 13.945(72)& 13.617(86) & 14.06&  0.206& 0.385& -0.24& -0.21& -0.09& 0.29\\
 \hline
\end{tabular}
\end{table*}

\begin{table*}
\centering
 \caption{Infrared magnitudes for program stars. 
Phases refers to ephemeris given in Table 1.}
 \begin{tabular}{@{}cccccccc@{}}
  \hline 
Object &I &J& H& K& JD/Date &phase &Source \\
\hline
\va~  &13.701(9) &13.346(22) &             -&12.954(116) &1998-08-12  &0.72   & DENIS Catalogue \\ 
\va~  &13.847(30)&13.472(90) &             -&13.050(180) &2450414.6148&0.48   & DENIS Consortium \\
\va~  &13.781(40)&13.560(110)&             -&13.134(160) &2451048.7763&0.81   & DENIS Consortium \\
\va~  &-         &13.545(42) &13.341(50)    &13.275(40)  &2451034.7109&0.69   & 2MASS\\ 
\vb~  &13.655(30)&13.388(90) &             -&13.316(210) &2450418.5524&0.42& DENIS Consortium \\ 
\vb~  &13.616(30)&13.383(130)&             -&13.079(150) &2451039.7991&0.79& DENIS Consortium \\
\vb~  &-         &13.403(29)&13.236(34)     &13.054(35)  &2451034.7134&0.77& 2MASS\\
\hline
 \hline
\end{tabular}
\end{table*}

\subsection{Spectroscopy}

\subsubsection{Optical lines and radial velocities}

We have identified many absorption lines in our spectra by correlating 
their central  wavelengths (measured by simple or multiple gaussian fits 
or simply with the cursor in a high resolution display) with vacuum 
atomic wavelength lists. 
For most of these lines we have obtained radial velocity measurements 
which are summarized in Table 5. 
We find only small variances among the velocities of the various ions,
at least for the primary components of lines formed in 
those ions prevalent in the circumstellar gas cooler than 10\,kK.
The average for 178 
lines in \va~ is 107 $\pm$ 12 km\,s$^{-1}$  and  for 75 lines in \vb~  
is 105 $\pm$ 9 km\,s$^{-1}$ (2007 data). This figure is close
to the same for the 2001 spectra  and indicates almost
the same radial velocities for the circumstellar regions of both stars.  
We observe moderately strong $V$ and $R$ emission components 
in the lower members of both the Balmer and Paschen series. 
The relative strengths of these features decrease to invisibility
by H$_{\zeta}$. 
These lines are flanked by absorption wings, typically stronger in the
blue wing, and have a deep central absorption.
The central absorption cores are found blueward the centroid of the $V$ 
and $R$ emission peaks typically by a few km\,s$^{-1}$ (Table 6). 
Remarkably, we  found blue absorption components (BACs, discussed later)  
in most metallic lines of \vb~ with a blue shift  with respect to the primary absorption line components.
This shift averaged -50 km\,s$^{-1}$ in 2001 and -58 km\,s$^{-1}$ in 2007.
The differences between these two epochs while small are probably significant.

\begin{table}
\centering
 \caption{Summary of heliocentric radial velocities (in km/s)  for 
MIKE spectra.  For \vb~ we give the velocity of the red (assumed 
``main") component and the velocity of the BACs when available after the
semicolon. The number of lines included in the averages is also listed.}
 \begin{tabular}{@{}ccccccccc@{}}
  \hline 
Ion & \va~&\vb~ &Ion & \va~&\vb~& Ion & \va~&\vb~\\ 
\hline
CaI  &108$\pm$12 (2)  &- ; 56 (1) &HI &106$\pm$3 (28) & 105$\pm$6 (28) ; -&ScI &112 (1) &- ; 58 (1) \\
CaII  &106$\pm$5 (5) &- &MgI  &103$\pm$6 (3)&104 (1) ; - & ScII &115$\pm$14 (2) &- \\
CrI & 81 (1) &- & MgII &115$\pm$24 (3) &141 (1) ; -  &SiII  &127$\pm$27 (7) & -\\
CrII &104$\pm$11 (10) &100$\pm$10 (3) ; 54$\pm$11 (6)  &NI &105 (1) & -&SrII &106$\pm$13 (2) &- \\
FeI  &110$\pm$12 (27) &- ; 49$\pm$3 (2)  &NaI &109$\pm$2 (2) & 108$\pm$3 (2) ; - &TiI  &98$\pm$26 (7) &- \\
FeII &106$\pm$3 (41) &104$\pm$6 (30) ; 53$\pm$5 (21)  &OI &104 (1) &- & TiII &105$\pm$7 (35) &110$\pm$16 (9) ; 51$\pm$7 (6) \\
\hline
\end{tabular}
\end{table}

\begin{table}
\centering
 \caption{Heliocentric radial velocities (in km/s)  for H\,I emission line components and shift of the central absorption relative to the centroid of the red and violet emission peaks. 
NP means not present. Note the larger peak separation in higher order lines.}
 \begin{tabular}{@{}ccccccc@{}}
  \hline 
Emission line &star &year & blue-peak &central-abs &red-peak&shift \\ 
\hline
H$_{\delta}$ &\va&2007&        20&     107&   172 &-2\\
H$_{\gamma}$&\va&2007&    31&     93&   168&-6\\
H$_{\beta}$&\va& 2007&           32&     92&   166&-7\\
H$_{\alpha}$&\va&   2007&       46&    91&   152&-8\\
H$_{\delta}$ &\va&2001&        -&     -&   -&-\\
H$_{\gamma}$&\va&2001&    -&     -&   -&-\\
H$_{\beta}$&\va& 2001&           24&     95&   159&-4\\
H$_{\alpha}$&\va&   2001&       55  &    93&   153&-11\\
H$_{\delta}$ &\vb&2007&        NP&     104&   NP&-\\
H$_{\gamma}$&\vb&2007&    NP&     106&   NP&-\\
H$_{\beta}$&\vb& 2007&           NP&     112&   181&-\\
H$_{\alpha}$&\vb&   2007&       64&     111&   165&-4\\
H$_{\delta}$ &\vb&2001&        -&     -&   -&-\\
H$_{\gamma}$&\vb&2001&    -&     -&   -&-\\
H$_{\beta}$&\vb& 2001&           46&    105 &195      &-16\\
H$_{\alpha}$&\vb&   2001&       56&     114&   170&-1\\
\hline
\end{tabular}
\end{table}

\begin{table}
\centering
 \caption{Equivalent widths ($EW$), maximum intensity relative to the continuum ($I/I_{c}$) and 
 full width at half maximum ($FWHM$) for the H$_{\alpha}$ emission line in 2001 and 2007.}
 \begin{tabular}{@{}cccc@{}}
  \hline 
Object & $EW$ (\AA) & $I/I_{c}$ & $FWHM$  (\AA)   \\
\hline
\va~  & 112-104 &24.1-22.7  &4.39-4.46  \\
\vb~  & 28-26   &6.3-5.7    &5.16-5.25  \\
\hline
\end{tabular}
\end{table}

\subsubsection{Photospheric lines, T$_{\rm eff}$, 
and FUV radial velocities of \va~ and \vb}

  For \va~ we estimated a spectral type of O9-B0, based, first, on 
fitting of the wings of both He\,I 3821\,\AA~ and H${\beta}$ as 
compared to these wings in O7-B0 Ib-III stars taken from the UVES POP 
webpages (Bagnulo et al. 2003). The (usually red)
wings of the Balmer and Paschen lines were also used to confirm 
this spectral type and Teff, given the implied luminosity class of III.
The C\,III 1176\,\AA~ widths/wings suggest a slightly cooler spectral 
type of 30-32\,kK. Based on this, we used 31\,kK, log\,g= 3.5 
from Kurucz models (Kurucz 1993). The precise photospheric
temperature does not appear to be important in the modeling 
of the CS absorptions discussed herein.

It is not clear how many lines in our spectra of \vb~ are photospheric, 
but they are few in number. The best candidates are the far wings of the 
Balmer and Paschen lines and the wings of C\,III 1176 \AA~ 
(Figs.\,4 and 5).  From the similarity in the $V$ magnitude we also 
take this star to have a luminosity class of about III. 
The wings of the hydrogen
lines are more developed in \vb, suggesting a slightly later type. However,
the C\,III 1176\,\AA~ feature is indistinguishable from the \va~ line.  
To split the difference between the implied values of $T_{eff}$, we used 
a Kurucz model of 29 kK, log\,g= 3.5 to model the photospheric fluxes.  

  Because of the dearth of photospheric lines we have had to rely on
the radial velocity differences of the Galactic (LISM) and local (SMC)
ISM line components. We then tied the LISM radial velocity to the Sun's 
so that this ISM velocity difference could be taken as an estimate of the
program stars' systemic velocities.  ISM components of a number of low 
excitation ions are plentiful in our {\it FUSE} spectra, and we were able 
to measure velocity differences for both stars.  Our results for \va~ 
and \vb, respectively, were 119${\pm 4.9}$ km\,s$^{-1}$ (39 lines) and 
110${\pm 6.1}$ km\,s$^{-1}$ (25 lines), referencing the zeropoint to the 
measured LISM wavelengths.
As a check on these results, we 
consulted the Tomlinson et al.  (2002) compendium of measured SMC ISM 
lines. The comparable-brightness objects lying closest in the sky to our 
program stars, AV\,14 and AV\,15, have RVs of 124 km\,s$^{-1}$ and 
130 km\,s$^{-1}$ (with ${\pm 9}$ km\,s$^{-1}$), 
which are in the expected rough agreement. 
Furthermore, the measurements of the optical lines discussed in section 4.2.1
are consistent with these RV values 
and a low velocity CS expanding medium in both our 2001 and 2007 data.

\subsubsection{The circumstellar features in \va}

{\it a. General description:}

With the exception of the broad wings of $H{\beta}$, He\,I 3821\,\AA, 
and possibly of the far-UV C\,III 1176\,\AA~ multiplet, none of its 
optical or near-IR lines are photospheric. Rather, they reflect different 
kinematic and thermal properties of an extensive circumstellar envelope. 
The orbit of a dense property may be 
responsible for the near or strict periodicity with a period of 119 days. 
The high-level Balmer and Paschen lines can be resolved to H30
and P24, respectively, leading us to estimates of a mean envelope volume 
density of $\sim$ 1$\times$10$^{11-12}$ cm$^{-3}$. 
The first few members of the Balmer line spectrum show strong emission, 
with $I_{H\alpha}$ = 24$I_{c}$, and $V/R$ = 0.9 (Table 7 and Fig.\,6).
We have no direct information on the degree of flattening of this disk
structure, but the separated $V/R$ H$_{\alpha}$ components suggest 
a quasi-Keplerian rotation of this structure.  We say ``quasi"-Keplerian
because of the likely very slow outflow of the gas from the star discussed 
below. Hereinafter, we will refer to a slow wind and quasi-Keplerian disk
interchangeably.

  The optical spectrum of \va~ exhibits only various permitted 
lines formed at ``hot," ``warm," and ``cool" temperatures. 
These lines are mainly in absorption, but a few Fe\,II lines 
exhibiting emission wing components will be highlighted below. 
The high excitation lines of He\,I, 5876 \AA~ and 7865 \AA, 
are present weakly in emission. Their profiles are narrow
 and they 
can be fit with {\sc circus} models having (``hot") temperature of 
15--25\,kK (probably closer to the upper limit) with column densities of 
5-10$\times$10$^{22}$\,cm$^{-2}$. The principal metallic line features
are Fe\,II and Si\,II absorptions, and these are formed at ``warm" 
temperatures of 6-9\,kK (Fig.\, 7).  We suspect also
that the excited O\,I 7771-5\,\AA~ triplet ($\chi$ = 9\,eV) is formed in 
the same warm region as the strong Fe\,II lines. Both groups of lines 
have turbulence-broadened wings and are formed optimally in a warm
environment. The resonance doublet of Na\,I and K\,I doublets 
are both formed at ``cool" ($\le$6\,kK) temperatures. 
Although the presence of these lines
give the red/near-IR spectrum the appearance of an F-G supergiant 
photosphere (M06), close inspection at high resolution demonstrates that 
the Na\,D lines have multiple components and a shaded blue-wing, suggesting 
a highly differentiated velocity structure. In our {\it FUSE} spectra,
we are able to make out both the Galactic and SMC ISM components; the
latter appear to show only one significant component toward either of our
program stars. The radial velocities (RV) of the Na\,I lines range from 
those exhibited by the Balmer, Fe, and Si lines to the largest 
measured, possibly indicating an outflow 
with a larger velocity gradient than 
found in the component forming the latter group of lines. 
Fig\,4 shows that in the case of \vb~ 
a portion of the line (blue wing) is formed in a region 
flowing outward faster than the flow of the mean disk.

One interesting aspect of the strongest metallic lines of \va~
(but not \vb) is that in addition to their sharp cores they also have 
strong wings (Fig.\,8a). However, these wings are neither extensively
tapered nor symmetric, i.e., the profiles do not resemble Lorenzians. 
Also, because some of these wings, notably those of Fe\,II 5018\,\AA,~ 
5169\,\AA,~ and 5197\,\AA~ are in emission, and because the excitation 
potentials of the lines can be low, we strongly doubt that these 
lines components are photospheric. 
In the case of Fig.\,8a 
we obtained similar fits with microturbulence
($\approx$50 km\,s$^{-1}$) or macroturbulences ($\approx$60 km\,s$^{-1}$) 
and for column densities of at least twice those (i.e., 
$\approx$2$\times$10$^{23}$ cm$^{-2}$) needed for the inner 
region of these lines. 

{\it b. Fitting equivalent widths with~ {\sc circus}:}

Our modeling proceeded in most cases with a single microturbulence 
value ($\xi$ = 5-10 km\,s$^{-1}$) and one temperature. The sample of
fitted lines shown below is a strategically chosen subset of a larger 
group of fitted lines. Except for He\,I lines and Na\,I and K\,I doublets, 
we are able to model all the UVES lines of both \va~ and \vb~ with T$_{c}$ 
values of either 8\,kK or 5\,kK. In some short stretches of the 
spectrum two different temperatures are necessary (Fig.\,7), 
meaning that actually the temperatures vary smoothly with position from 
the star. We have also been able to fit the emissions of Fe\,II 5018\AA~ 
(Fig.\,8b)
using an LTE representation (emission + absorption) components. 
Other lines are represented by absorption only, i.e. the source function
in the line is zero. 
Excepting the hydrogen lines as one proceeds to less excited species, 
higher and higher column densities are 
necessary to fit the spectra, e.g. 1--3$\times$10$^{22}$ for He\,I, and 
about 1$\times$10$^{23}$ cm$^{-2}$  for the 5\,kK and 8\,kK gas 
components. In contrast, the optical depth of the high level Balmer lines,
which are formed in gas with all these temperatures, suggests 
a total hydrogen-absorbing column density of about 10$^{24}$ cm$^{-2}$
and a volumetric density of 10$^{11-12}$ cm$^{3}$.

  We have attempted to fit the $H_{\alpha}$ emission strength. 
Using our LTE models with $T_{c}$ = 9\,kK and $\xi$= 10 km/s, we found 
an implied total area of $\sim$10\,000 stellar areas, such that the 
extent of the disk is roughly $\sim$ 100$R_{\star}$. 
This is roughly consistent within the errors with the $V, R$ velocity
separations indicated in Fig.\,6 and also with the disk emission area
of 2\,500 areas we derive from the Fe\,II 5018\AA~ emission (Fig.\,8b).
However, we reiterate that these fittings were done with LTE models, 
the numerical value we derived especially for H$\alpha$ should be 
taken at most to an order of magnitude.

  Additional evidence of a warm emitting gas seen beyond-the-limb sight lines
is furnished by the weakened C\,III 1176\,\AA~ multiplet, which requires 
hot gas to mimic a $\approx$25\% dilution of the line by an additional 
radiation field. We require a $T_{c}$ = 25\,kK component having about 5
emitting stellar areas to accomplish this (Fig.\,5a).
In principle, these emission areas can be used to compute hydrogenic 
Stromgren spheres.  We estimate that the radius of this sphere is that 
of a O9.5-B0 star, consistent with our estimate in Section 4.2.1 from 
the 1176\,\AA~ and optical H\,I and He\,I lines. 
We also point out that the 25\% dilution for C\,III 1176\,\AA,~ 
suggesting a substantial far-UV continuum emission component, from a
volume {\it around} the star, is greater than the several percent 
dilution we observe in the D lines. This suggests that flux contribution
in optical wavelength is smaller, and this is generally consistent with
the inference that the far-UV continuum emission comes from a region cooler than the
photosphere.

\subsubsection{The circumstellar features in \vb}

  Nearly everything stated about \va~ in its introduction is also true of
\vb.~ Once again it was difficult to find photospheric absorption lines; the
\vb~ spectral type may be O9-B0 or B0, according to its stronger H-line wings. 
The high level Balmer line limits are H29 and P23 (compared to H30 and 
P24 for \va), so 
its mean volume density and total column density is close to than the disk 
densities for \va.~ The $H_{\alpha}$ emission is 4 times weaker than for 
\vb~ ($I_{H\alpha}$ = 6$I_{c}$), and V/R $\sim$ 1 (Fig.\,6). 
Assuming a similar excitation temperature and a similar electron density 
from the visibility of the high level Balmer lines, the gas component 
of the \vb~ disk is about half as extensive as the \va~ disk.

  Multiple temperature fits were constructed using {\sc circus} for 
several features in the UVES spectrum, just as with for \va.~ 
The gas in \vb's disk likewise shows layered temperatures.
Again as with \va,~ there is evidence of line or continuum
emission close to the star, and  there is a very faint 
emission in the He\,I 5015 \AA~ line. The 5876\,\AA~ and 7065\,\AA~ 
lines consist of both weak emission and absorption components. Also,
our remarks for the dilution of the \va~  C\,III 1176\,\AA~are equally
valid for the fitting of this feature in the \vb~  spectrum - see Fig.\,5b.
As before, in order to fit the C\,III\,1176\AA~
feature we had to introduce a hot emitting region of about 5 stellar areas
to the photospheric flux.  For both stars it appear that this flux is
consistent with the appearance of the He\,I lines in the optical spectrum. 

  In the \vb~ spectrum the observed lines are formed exclusively in
absorption and arise from a diversty of excitation states.
These include the hydrogen lines, the high excitation O\,I 7771-5\,\AA~ 
triplet (9\,eV), the many lines of Fe\,II and Si\,II,  and finally the 
Na\,D and K\,I resonance doublets.  Our models show that
the K\,I line is formed at a temperature of 5\,kK or less. A comparative
weakness of the K\,I and O\,I lines in the \vb~ spectrum is 
consistent with the only moderate emission in H$\alpha$.

The primary difference that the SMC-SC4 spectrum exhibits with respect to 
SMC-SC3 is the presence of one or two Discrete Blue Absorption Components 
in almost all the metallic lines. We call these features "BACs." 
These are blueshifted by 50-58 km/s relative to their ``main" absorption
members. As an example, the strongest Fe\,II lines (Fig.\,9a) are
accompanied by two blue BAC features each separated by -30 km\,s$^{-1}$.
The middle absorption component of these is absent 
in any other lines in our spectra. 
An important difference between the 5018\,\AA~ (2.9\,eV) and its 
slightly stronger multiplet members 4923\,\AA~ and 5169\,\AA~ is that 
that the 5018\,\AA~ line exhibits a broad absorption wing. The red
wings of the other two lines reach the continuum abruptly and hint
at the presence of red emission. This trend continues as one goes 
to less strong Fe\,II lines, such as 5276\,\AA~  and especially 
5272\,\AA~  (see Fig.\,7; lower spectrum).
The latter are somewhat filled in by emission, according to the 
pure-absorption models that fit most of our other weaker Fe\,II line 
absorptions well.  Yet weaker but still fairly strong lines arise
from the same excitation potential;  Fig.\,10a shows this emission for
5169\,\AA~ and also in the red wing of the Mg\,I 5172\,\AA~ line. Tracing
these characteristics altogether, from the intermediate-strength Fe\,II 
lines to the lines of the more abundant ions Mg\,I and Si\,II, there is 
a trend that the lines with the highest optical depth exhibit the greatest 
tendency for the ``main" red component to be filled in by emission. 
This is the typical signature of P\,Cygni emission in an outflow with
only a small expansion velocity. The emission originates from scattering 
of line flux from a large annular region in the sky plane around the 
star (though primarily along the disk axis). This conclusion is required 
by the required coverage factor of many stellar-areas to model the 
metallic line emissions in our {\sc circus} models (see Figs.\,8b, 9b, 
10a). These models demonstrate that the emission is too large to be caused 
by gas heating along lines of sight directly to the star.

Although the BACs are a highlight of \vb~   optical spectrum,
they are not unique to this star. In a little noted paper,  
Heydari-Meydari (1990) reported that the metallic absorption lines of 
the sgB[e] star N82 for the most part consist of double components, 
the dominant member of which is blueshifted by about -46 km\,s$^{-1}$.
In addition, we notice a faint vestige of BACs at the same velocity
also seem to be present in the Mg\,II 4128\,\AA, 4132\,\AA~  doublet 
in the \va~ spectrum - see Fig.\,10b.
These similarities prompt the speculation that our program stars 
are part of a larger group - we call them ``bgBe stars." Such stars must 
have extensive and kinematically complicated (but slowly expanding) 
disks in the central plane in which either high
turbulence or colliding wind components are common. If so, they may 
form a morphological if not evolutionary bridge between sgB[e] and 
classical Be stars.

\subsubsection{Comparison of the UVES and MIKE optical spectra}

 As noted in $\S$4.2.1, the measured 
radial velocities are practically the same from our high resolution 
UVES and MIKE spectra obtained in 2001 and 2007. The
roughly constant wavelength spacing between the absorption components
in Na\,D (one is the Galactic ISM component) suggests that there is 
little or no change in RV in the reference frame of the outer regions of 
the CS disk between our two observing epochs. One of the conclusions from
the absence of clear RV changes is that short-period binarity is unlikely.  
Second, we see almost no difference in the high level hydrogen or 
metallic line strengths of \va~ between the 2001 and 2007 spectra. 
For \vb\,there are only subtle differences in the line profiles.
Among these are: (1) the H$\alpha$ emission is slightly 
weaker; the $V/R$ ratio, now $>$1, is reversed, 
but still is close to one, and (2) the multiple
components in the strongest Fe\,II and Na\,D lines are sharper in 
the 2007 spectra. Thus, while the details of the wind flow are slightly 
different, the general slow outflow characteristics are the same.
For \vb~ the new spectra show several blue Fe\,II lines that were
not so clear before, and the generally sharper BAC components 
bring out the emission. Two of these are Fe\,II 4585\,\AA~ and 4632\,\AA;
the ``main" (red) component shows very clear P\,Cygni-like structure, 
which we could barely discern in the 2001 spectra. 

The 2007 spectra also cover the blue region of the Ca\,II H/K lines,
which were not included in the 2001 UVES observations. This doublet exhibits
developed wings. While the line cores do not reach the zero flux level, like
the D lines they are deep enough to suggest that a putative cooler companion 
could not contribute much flux in the blue spectral region.  
As with the Na and K resonance lines, these profiles would account from the 
previous F spectral type for \va~ based on low resolution spectra (M06).
We do not believe these features are photospheric, but rather that most of 
the column density of the disk is concentrated far from the star where 
T$_{c}$ $\le$8\,kK.  For one thing, the CaII K line core of \vb~  exhibits
a double-lobed structure, suggestive of the BACs in the very lines that
flank it. This and the fact that narrow cores drop down to low fluxes
suggests that at least these parts of the lines are CS in origin. The line
wings are far too strong to be formed in the photosphere of a $\sim$B0-type 
star. 

\begin{figure}
\scalebox{1}[1]{\includegraphics[angle=90,width=8cm]{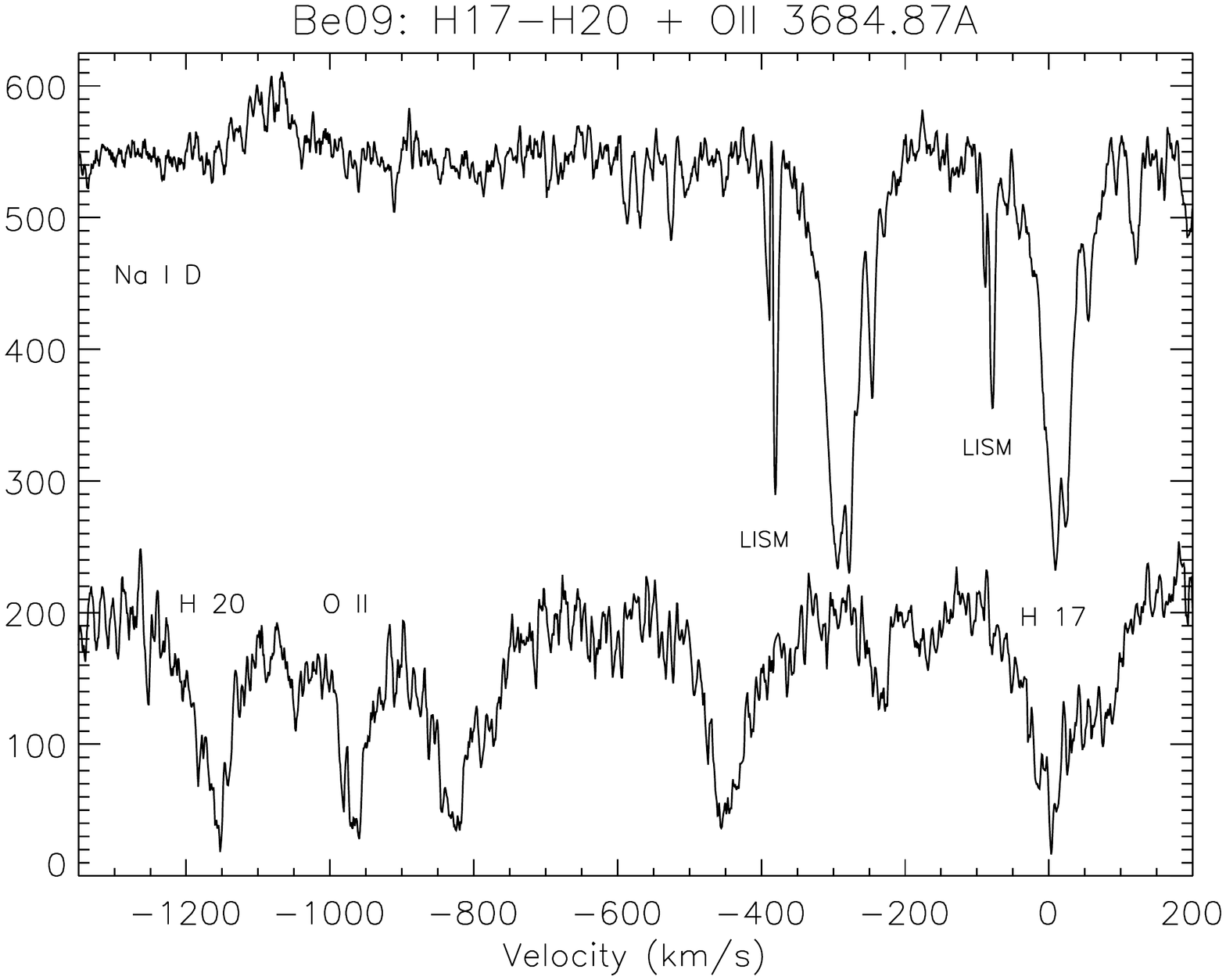}}
\scalebox{1}[1]{\includegraphics[angle=90,width=8cm]{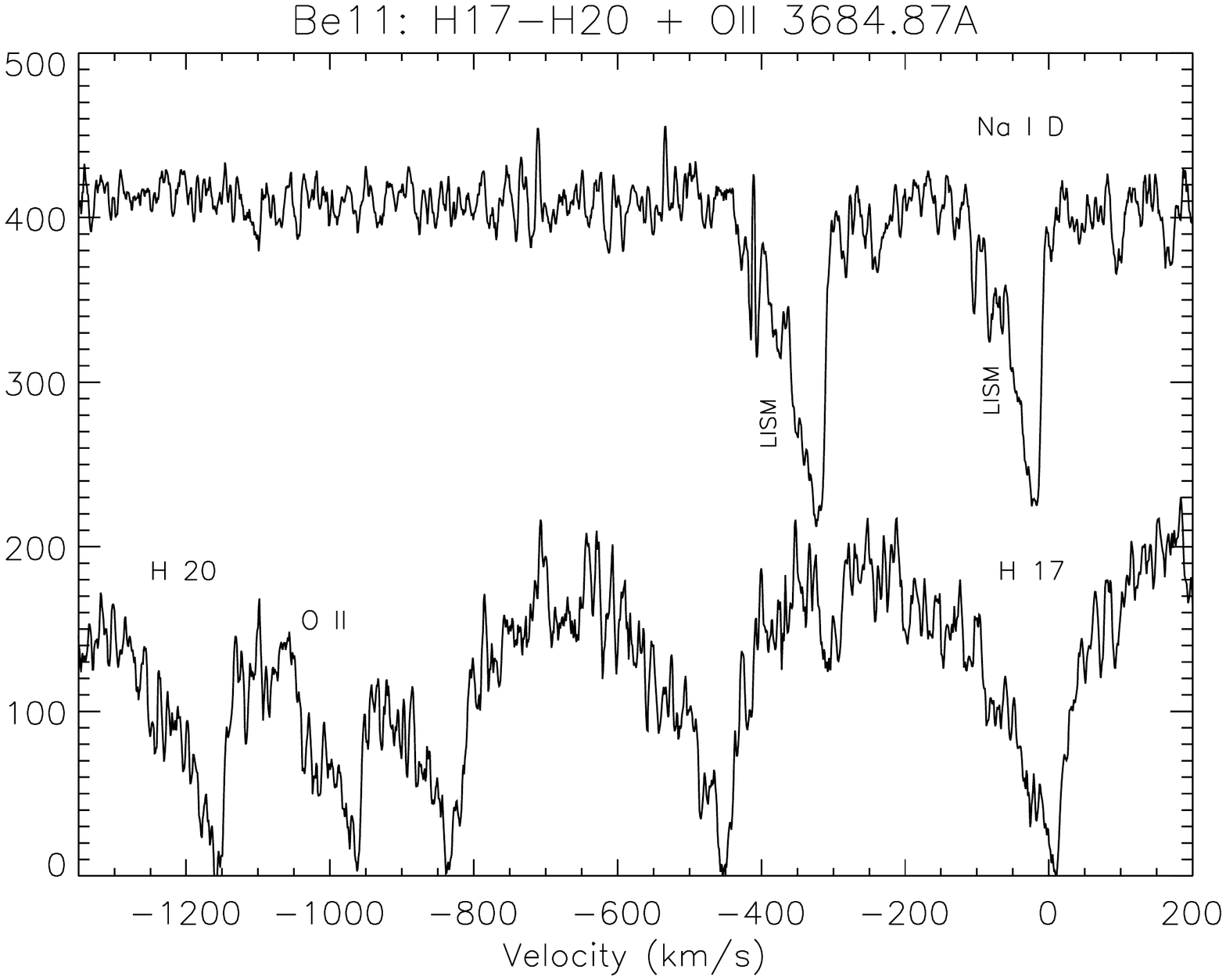}}
\caption{A comparison of the Na\,I D doublet and the high level Balmer 
lines in our UVES spectra for \va~ (left panel) and \vb~ (right panel).
The zeropoints of the velocity system are referred to the rest frames 
of the D2 (5896\,\AA~) and H17 lines.}
\label{4}
\end{figure}

\begin{figure}
\scalebox{1}[1]{\includegraphics[angle=90,width=8cm]{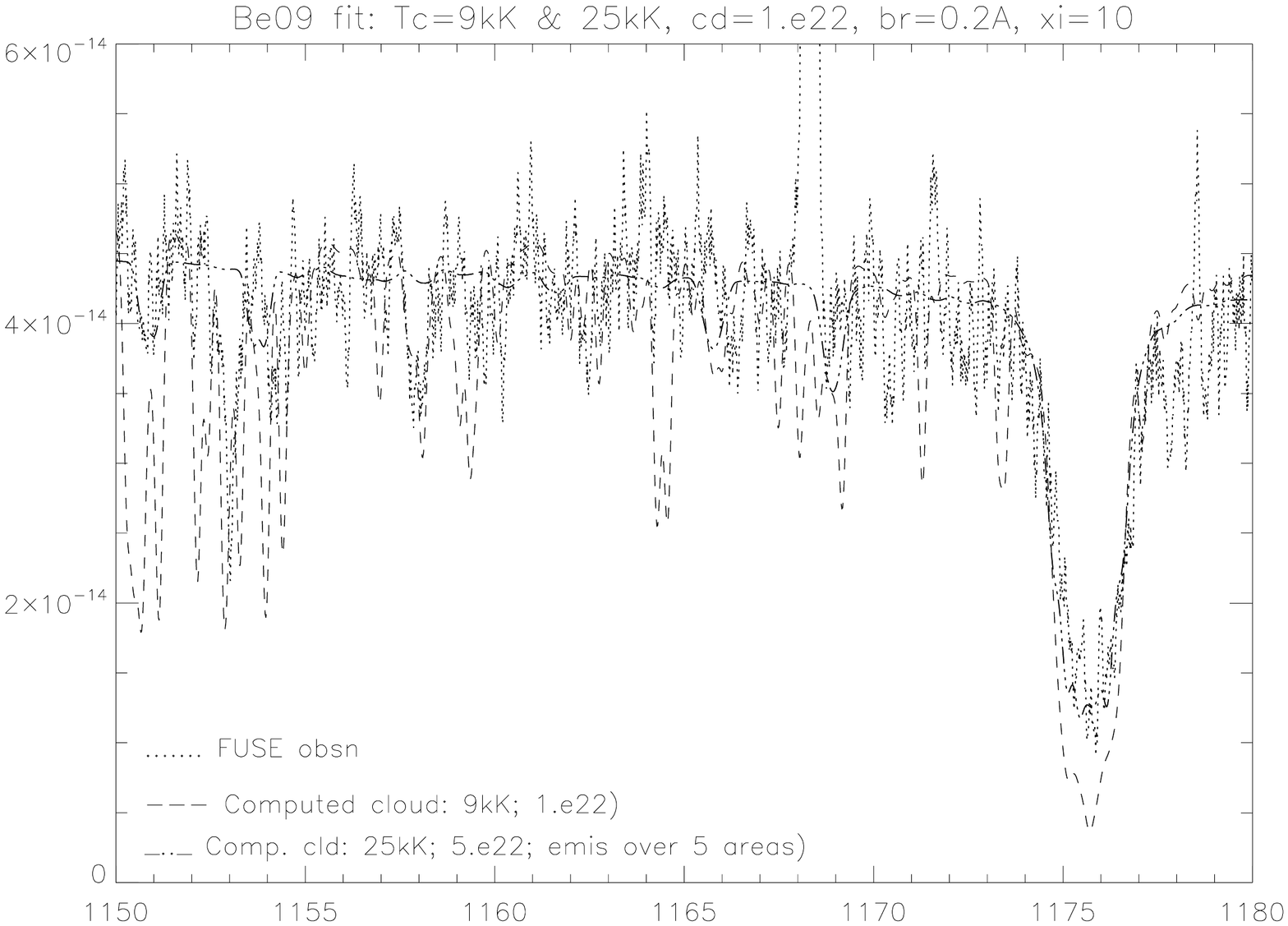}}
\scalebox{1}[1]{\includegraphics[angle=90,width=8cm]{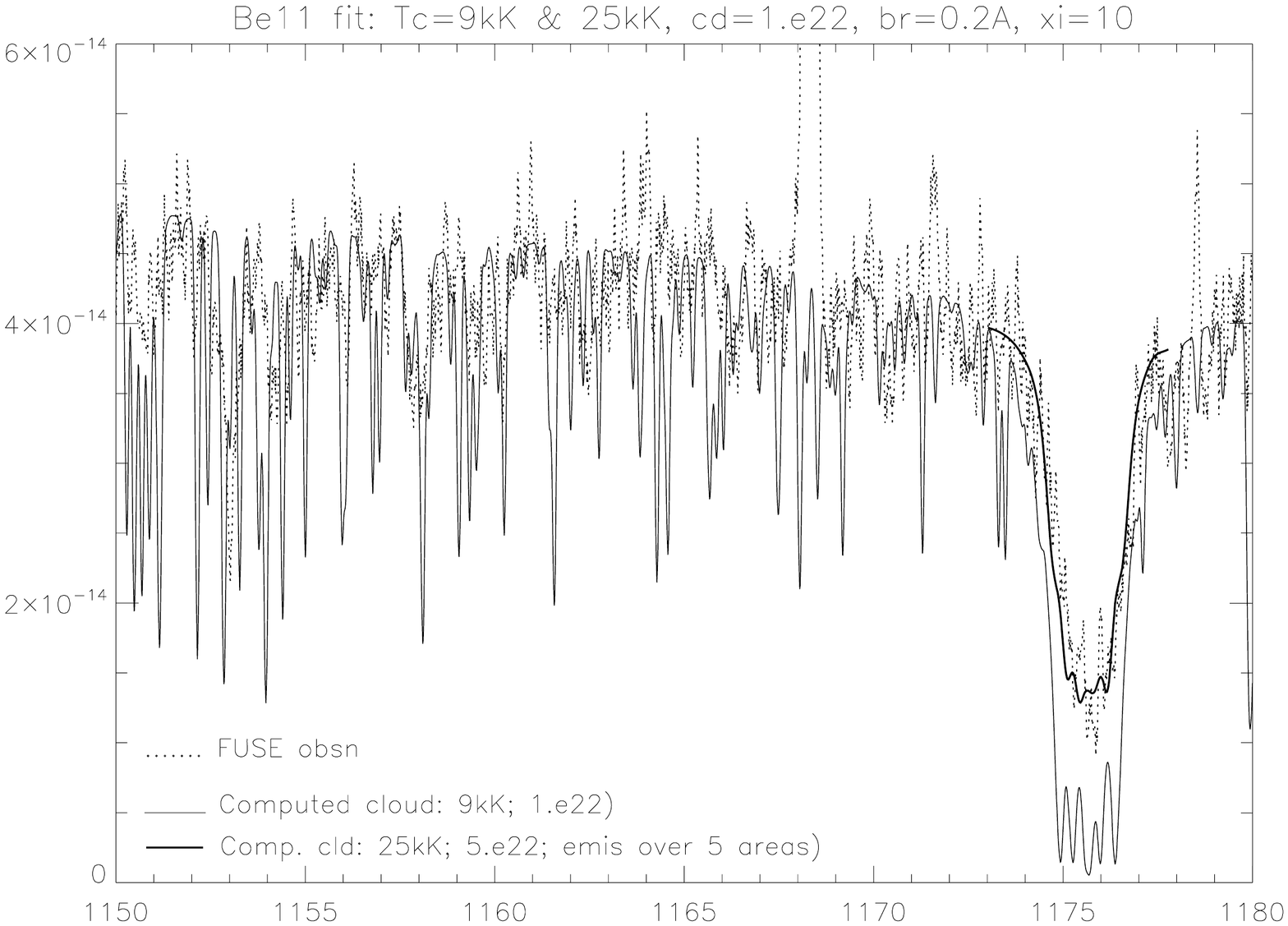}}
\caption{Fit of C\,III\,1176\,\AA~ for \va~ (left panel) and \vb~ (right 
panel). The dashed lines are FUSE observations, while the solid and 
dotted lines are fits without and with the emitting line dilution discussed 
in the text. Annotations indicate the model parameters.
}
\label{5}
\end{figure}

\begin{figure}
\scalebox{1}[1]{\includegraphics[angle=90,width=8cm]{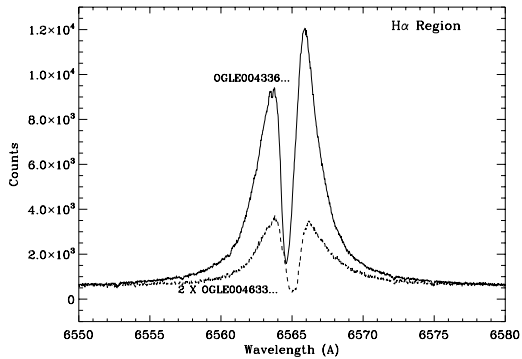}}   
\caption{The UVES spectra depicting the H$\alpha$ profile for both the 
program stars. }
\label{6}
\end{figure}

\begin{figure}
\scalebox{1}[1]{\includegraphics[angle=90,width=16cm]{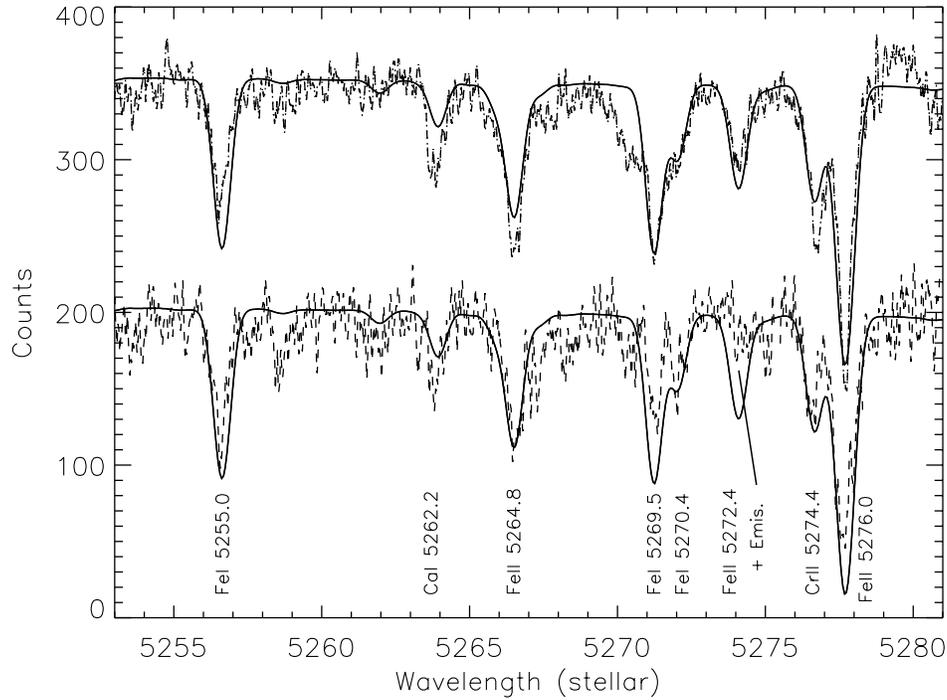}} 
\caption{ The two temperature fit of yellow Fe-like lines for \va~ 
(upper) and \vb~ (lower). In both cases temperatures of 8\,kK and 5\,kK 
were used and the warm column density was 1$\times$10$^{23}$\,cm$^{-2}$.
For the respective stars, the columns for the cool component were
1.5$\times$10$^{22}$\,cm$^{-2}$ and 0.5$\times$10$^{22}$\,cm$^{-2}$.
}
\label{7}
\end{figure}

\begin{figure}
\scalebox{1}[1]{\includegraphics[angle=90,width=8cm]{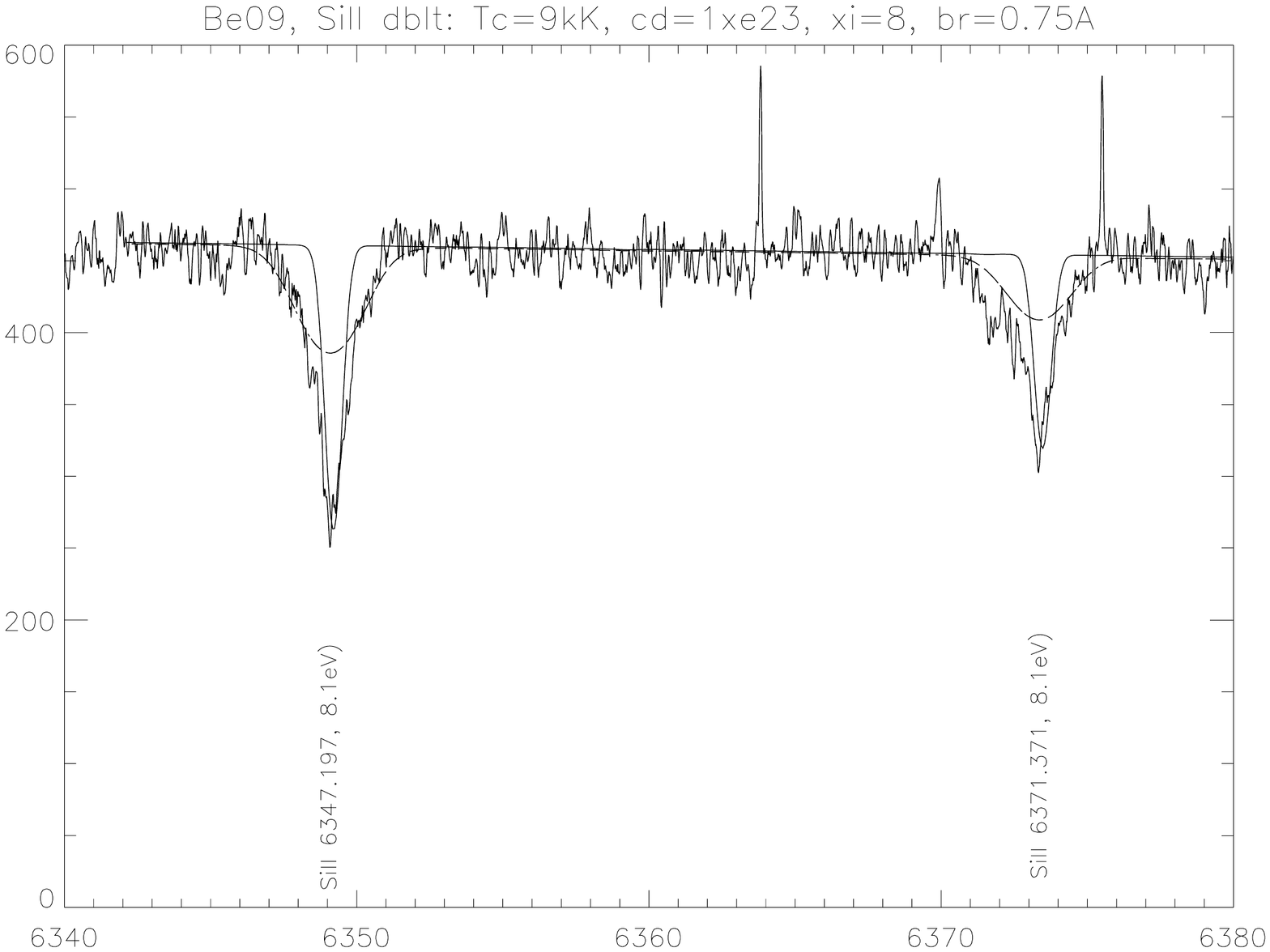}}
\scalebox{1}[1]{\includegraphics[angle=90,width=8cm]{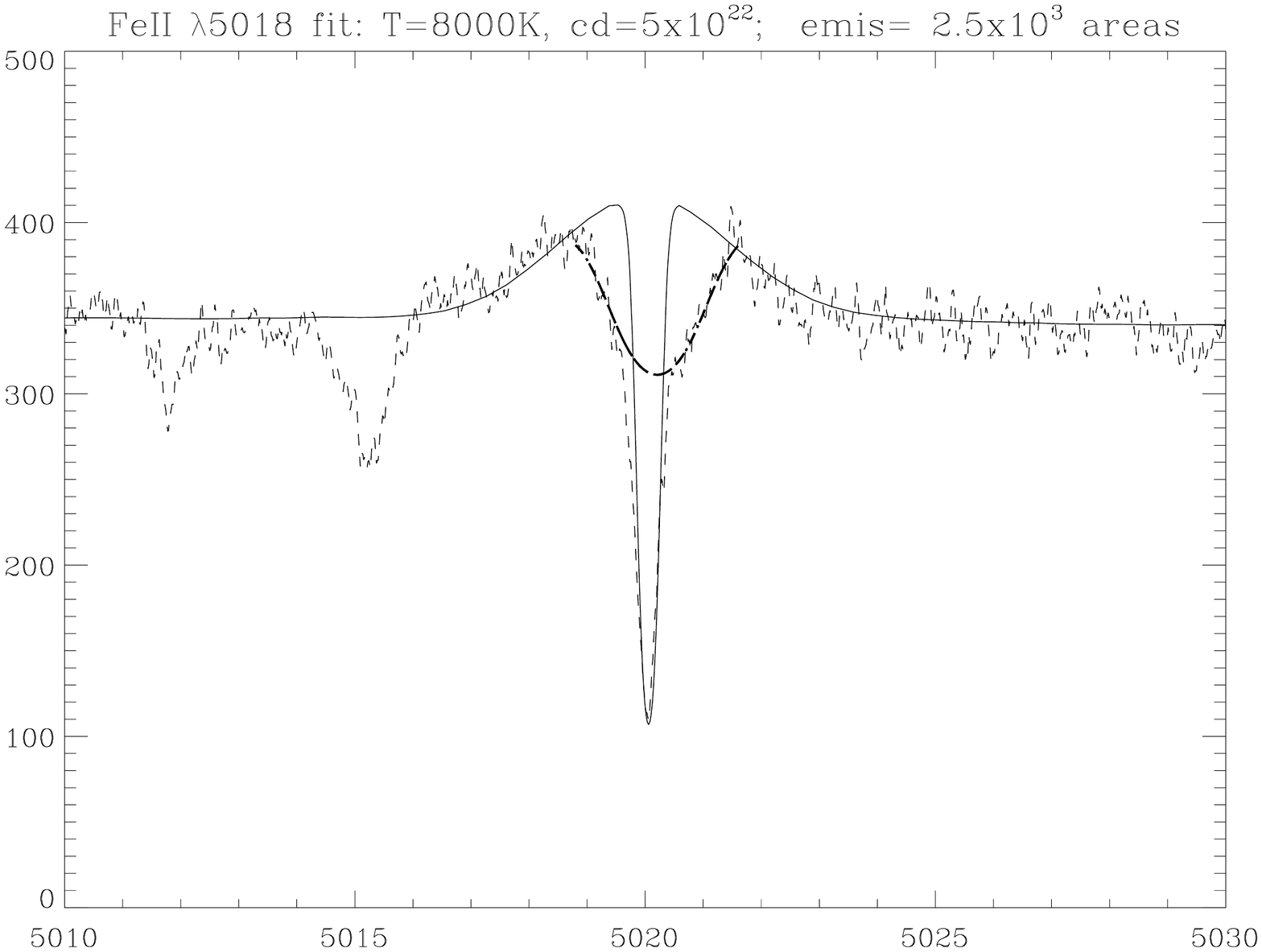}}    
\caption{
(a) A fit for the core and wings of Si\,II the 6347\,\AA~ and 6371\,\AA~ 
doublet for \va,~ using a temperature of 8\,kK and two separate turbulencies 
of 10\,km\,s$^{-1}$ and 50\,km\,s$^{-1}$ and column density fitting 
models.~ 
(b) A fit of the Fe\,II 5018\,\AA~ line for \va. This profile was
fit by two independent models, one with a temperature of 8\,kK and a 
column density of 5$\times$10$^{22}$\,cm$^{-2}$ and a low microturbulence
of 10\,km\,s$^{-1}$. The second component was fit to the emitting wings
with the same CS temperature, only with a medium extended over 
2\,500 emitting stellar areas and a  turbulence of 50\,km\,s$^{-1}$. 
}
\label{8}
\end{figure}

\begin{figure}
\scalebox{1}[1]{\includegraphics[angle=90,width=8cm]{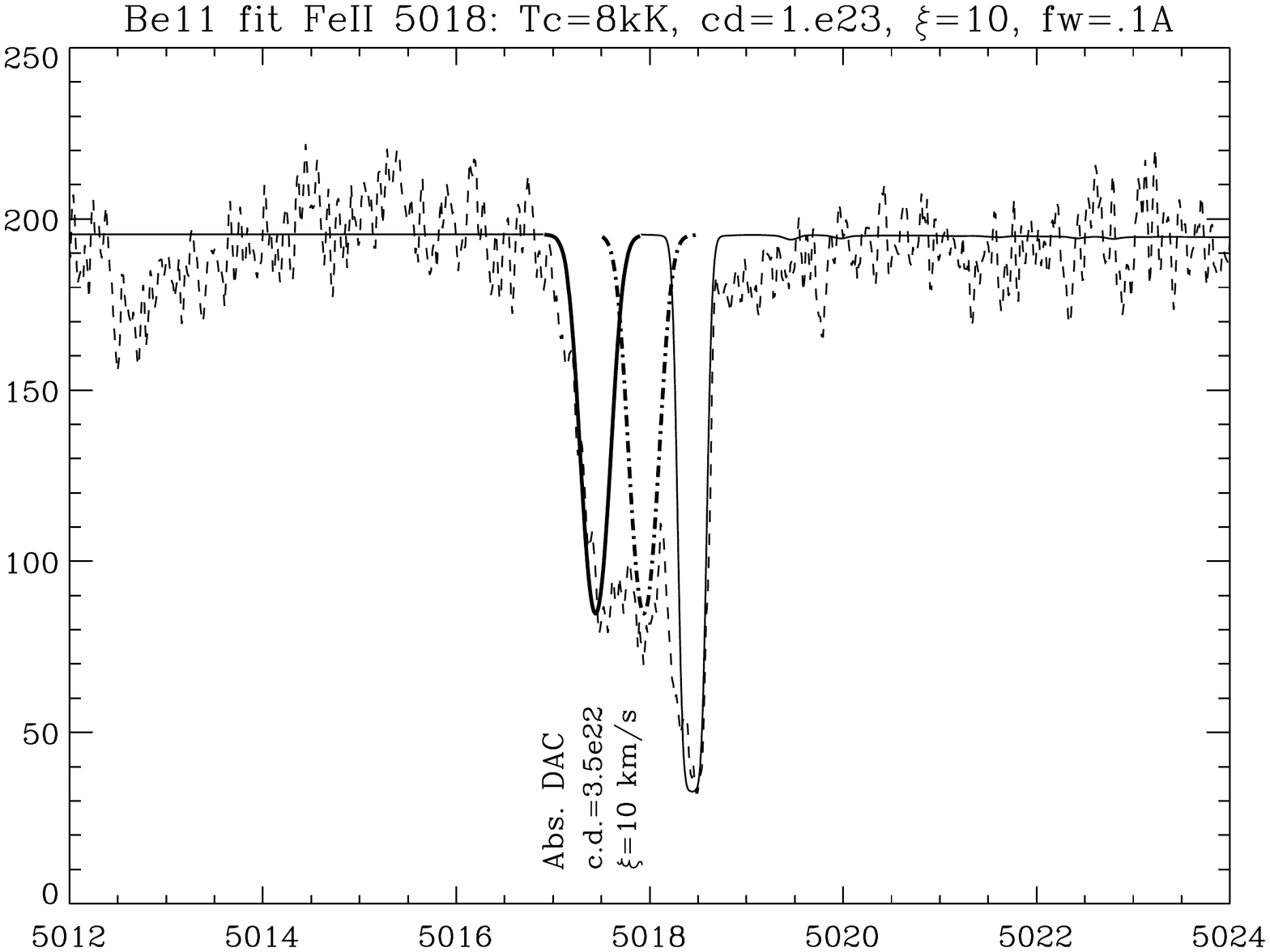}}
\scalebox{1}[1]{\includegraphics[angle=90,width=8cm]{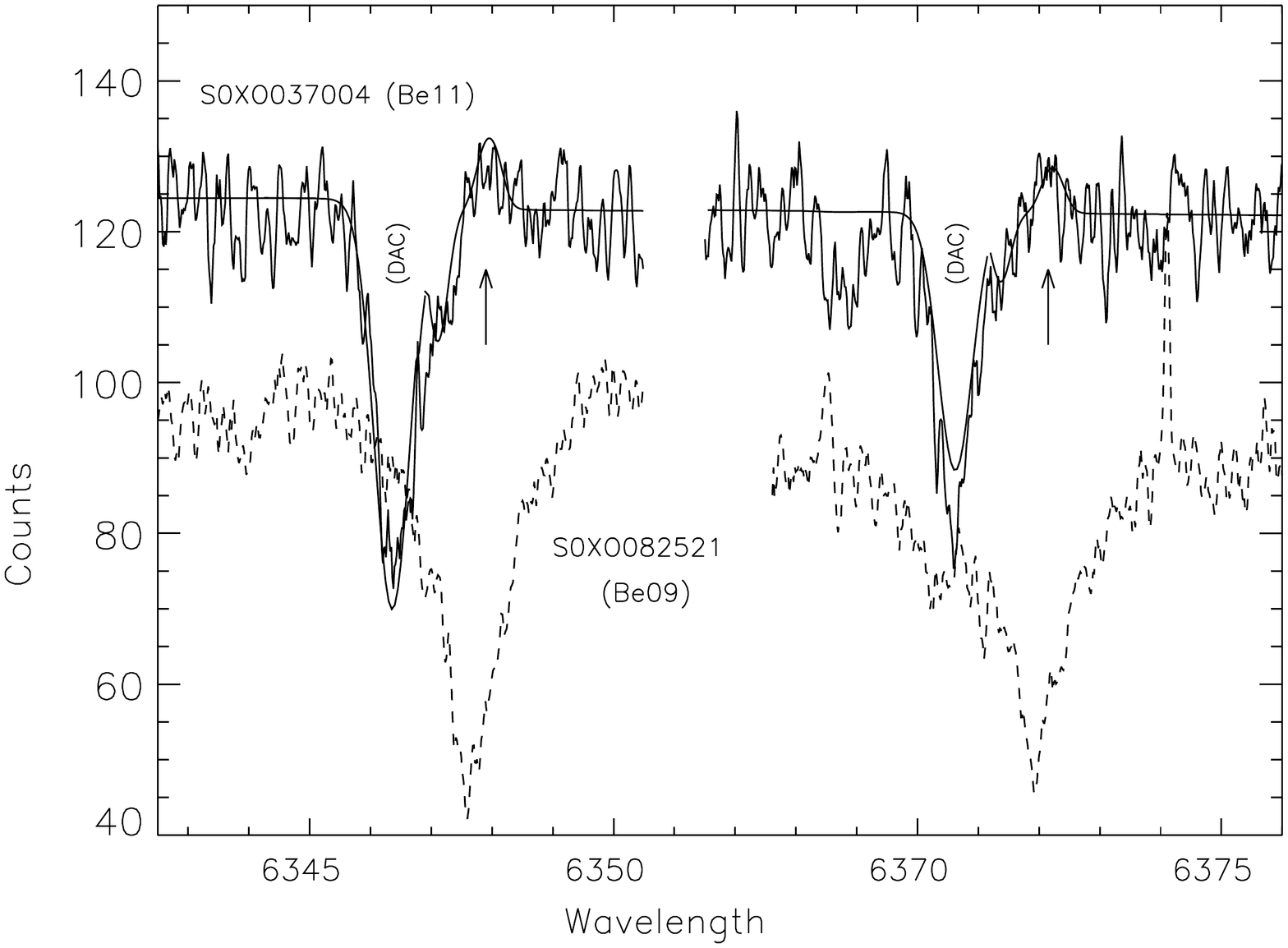}}   
\caption{(a) The Fe\,II 5018\,\AA line of \vb, exhibiting {\it two} 
Discrete Blue Absorption Components (``BACs").~
(b) A fit to the Si\,II  6347\,\AA~ and 6371\,\AA~ doublet (solid 
line), emphasizing the BACs and the filled in main components for the \vb.~ 
The observed spectrum (only) is shown for \va,~ in order
to guide the eye to the filled in emission in the \vb~ spectrum. For both 
spectra the 6371\,\AA~ line has been displaced for convenient reference.}
\label{9}
\end{figure}

\begin{figure}
\scalebox{1}[1]{\includegraphics[angle=90,width=8cm]{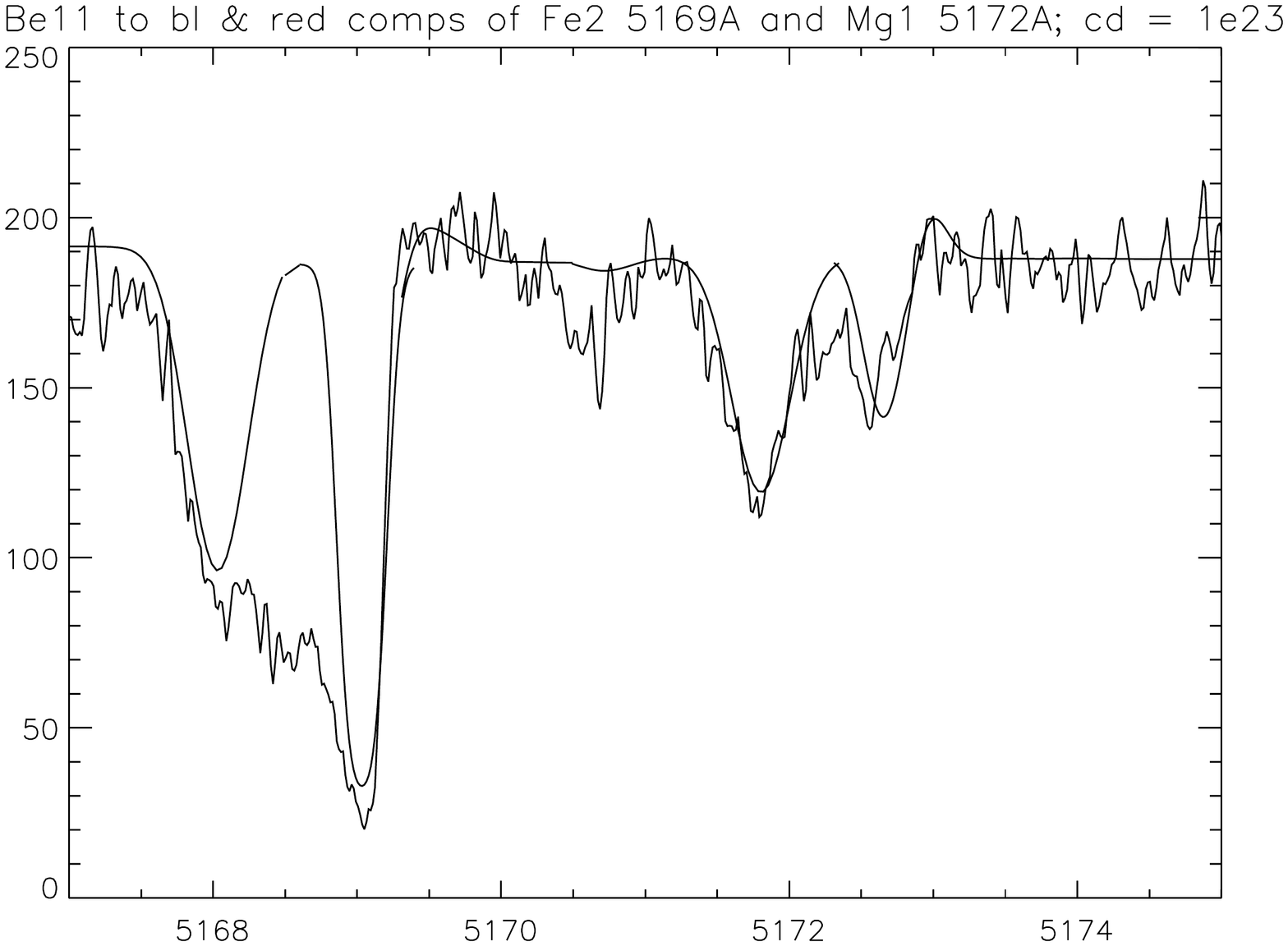}}    
\scalebox{1}[1]{\includegraphics[angle=90,width=8cm]{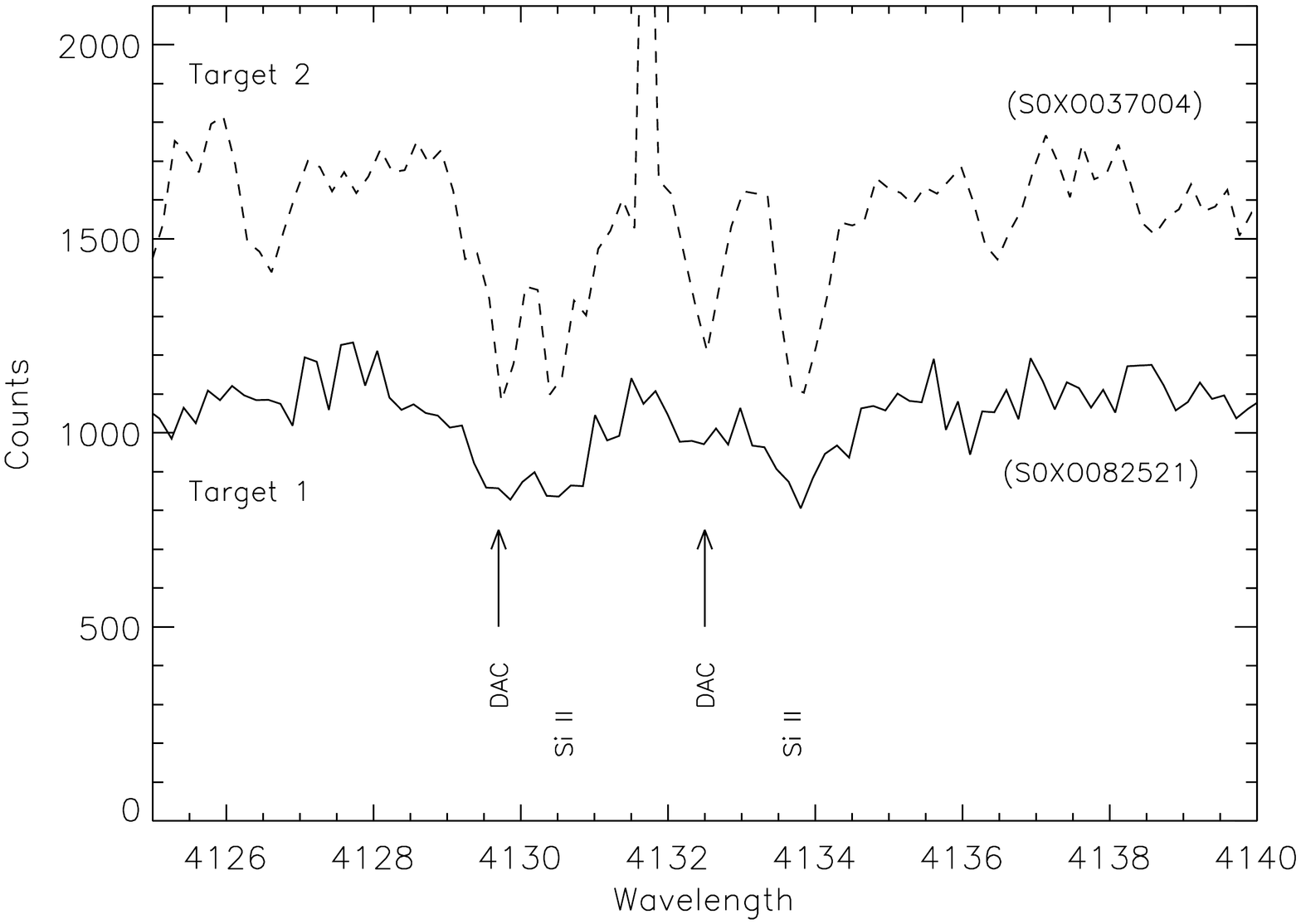} }
\caption{(a) The region around the Fe\,II 5169\,\AA~ and Mg\,I 5172\,\AA~
lines in the spectrum of \vb.~  The observations were fit to a two 
absorption model, with the BACs each being fit with a 8\,kK gas with
a column of 1$\times$10$^{23}$\,cm$^{-2}$ and blue shifted by -35\,km$^{-1}$;
the red component was fit with a column of 3$\times$10$^{23}$\,cm$^{-2}$.
The Mg\,I emission component was produced with 10 stellar areas having 
this same temperature; the ``main" (red) component of the Fe\,II line 
was unaffected by this emitting area.  
~(b) A comparison of the profiles of the 8\,eV Si\,II 4128-4132\,\AA~
doublet for \vb~ (dashed) and \va. Discrete blue absorption components, 
while clearly seen in the \vb~ spectrum, are barely visible in \va. 
}
\label{10}
\end{figure}

\section{Discussion}

In order to interpret the variability and spectral characteristics 
observed in our targets, in principle we can invoke mass loss/exchange 
in a binary system with two nearly equal mass components. 
However, we have rejected this hypothesis because of the unmeasurable 
changes in RV,  small departures from strictly periodic variability in the 
light curve, and the absence of a secondary component in the hydrogen
and resonance line spectra.
Rather, to date the evidence supports the view that these objects are 
effectively single stars surrounded by a slowly expanding disk which is 
fed by a moderate velocity wind close to the star.

The chances of our intitially selecting two stars with unexpectedly 
anomalous spectra out of a group of eight otherwise normal B stars are 
very low.  Therefore, we suspect that the
bgBe stars are comparatively numerous in the SMC fields S02 surveyed, 
perhaps accounting for at least 10\% of all the Type\,3 B variables 
down to $V$ = 14.
Even so, somehow these objects certainly represent a short-lived stage 
in the evolution of OB type stars in at least the SMC. 

The most striking spectroscopic characteristic of these objects is their
confined thick disks, which seem to be tied to winds. Although we 
do not yet know whether, like many B[e] stars, these CS systems have a 
hybrid nature, it is tempting from our observations to speculate that 
the wind has both a slow and moderate or fast velocity 
component along the same sight lines.  In this context, Madura et al. 
(2007) have discussed B[e] wind models in which a fast wind solution 
abruptly develops a ``kink," quickly transitioning to a slow component. 
The kink solution is due in part to the rapid rotation of
the star.  Rapid rotation could have been initiated either after pre-main
sequence evolution (an oft-advanced picture for classical Be stars), from
spin-ups via intra-binary mass transfer from a now evolved companion (like 
$\phi$\,Per), or from spin-up during a blue loop phase of late (single-star) 
stellar evolution, as sometimes posited for sgB[e] variables (Langer 
\& Heger 1998). The evidence does not rule out any of the three scenarios, 
although the case for mass transfer in a massive close binary
(i.e., with an orbital radius less than than a few A.U.) 
so far seems the weakest.

Regardless of how the bgBe stars evolve to the early B (roughly) class 
II-III region of the HR Diagram, fundamental questions remain concerning 
a clearly complicated geometrical and kinematical structure.
In global terms, the disk must be an equatorially confined disk
with a possible, nonaxisymmetric component, which undergoes Keplerian
rotation and a small outward outflow component near this plane. 
In this last sense, the ``disk" seems to be closely tied to a slow wind, 
probably even slower in velocity than the slow equatorial winds of 
sgB[e] stars. In this one sense, the disk has intermediate kinematic
properties between the Keplerian disks of Be stars and slowly expanding
disk/winds of sgB[e] variables.

 We may summarize the evidence that the structure around the bgBe
stars in this study are disks from the following observations:

\begin{itemize} 

\item The double-peaked $V$, $R$ emissions of the peak in H$_{\alpha}$ 
  and other lower members of the lower Balmer sequence. This is a key
  signature of Keplerian rotation in equatorially confined disks, as seen
  by an observer viewing the system at an intermediate or edge-on angle.

\item The reversal of the $V$, $R$ emissions on a timescale of a few 
   years is likewise associated with Be-like disks. 

\item The half-power emission peaks of the lines are broader as one 
 moves up Balmer lines from H$_{\alpha}$. This too is the hallmark of
 Keplerian rotation in Be stars. The larger emission widths result
 from a lower opacity in the high level lines, thereby sampling a
 larger equatorial rate closer to the star.

 \item The presence of emission centered nearly in the wavelength frame 
  of the primary CS absorptions is consistent with Keplerian motion.
  
 \item The nearly periodic variations of the light curve strongly 
  suggest recurrent occultations by a structure in the plane. 

\end{itemize} 

  At the same time, there is evidence that the disk is actually part
of a very slow equatorial wind or outflow:

\begin{enumerate}

  \item strong lines such as the wings of the Balmer and Paschen lines,
  and in addition the Na\,D lines, exhibit a slight strengthening of the
  blue wing, which is consistent with an accelerating flow, at least 
  in some sight lines to the star.

  \item the cores of the H$_{\alpha}$ are shifted to the blue by 7-8
   km\,s$^{-1}$ relative to  H$_{\delta}$ (Table\,6). This puts a
   strong lower limit to an outflow velocity to the more superficial
   disk formation sites for H$_{\alpha}$ relative to the other lines.

  \item In the metallic lines of \vb,~ the redshifted emission in the
  Si\,II 6347\,\AA~ and 6371\,\AA~ doublet, some strong Fe\,II and 
  Mg\,I lines demonstrates a small blueshift in  the ``main" absorption 
  component.~

\end{enumerate}

  In fact, the description of these CS structures are  much more
complicated than merely of a slowly outflowing axisymmetric disk.  
Most of \vb's spectral lines exhibit one or even two ``BACs," which
beg an interpretation of collisions of a comparatively rapid outflowing 
gas with a more leisurely expanding disk.  Indeed, the radial velocities
of the BAC components are similar to those of the He\,I lines, suggesting
this component accelerates freely from the star's equatorial regions
over a short distance (probably $\le$R$_{*}$) before encountering 
the disk. We speculate this outflow maintains the disk mass as it expands
from the star. From Fig.\,10b, it appears that BACs are present in
the \va~ spectrum too. Their weaker appearance could be caused by
a number of factors, among them simply a less massive flow. However, 
we suspect the primary factor is that the inner flow is less visible 
if our viewing angle to this star is more edge-on and we cannot see
the inner disk interactions as clearly. This would be in agreement with
the smaller H$_{\alpha}$ emission, which is a function of projected
area, the larger $E(B-V)$, and also the similar IR re-emission (from 
optically thin gas and/or warm dust) with respect to \va.~

  The broad symmetrical emission or absorption wings in the spectral
lines of \va~ require a separate description - indeed one that cannot be
said to be unique because there are many ways of generating turbulence.
However, one can gain a start in understanding this by noting that
a turbulence of at least 50 km\,s$^{-1}$ hints at a source
with a considerably higher kinetic energy. The most plausible source
is of a high-velocity wind, which abrades the central disk plane as
it flows across it. This explanation would likewise explain symmetrical
emission due to scattered flux from gas from beyond the stars' limbs.
This explanation carries the prediction that a high-velocity wind 
is present. In this event, tapered blue wings should be visible in the 
wind-sensitive resonance lines in the ultraviolet. We would further
predict that the strengthening of these wings would be slight because
lines of sight only intersecting the most visible ``upper" limb of
the star would sample the high latitude wind.

   This latter point brings up the important point that in the
envisaged disk geometry, including the Kraus \& Lamers opening disks,
any observer at equatorial or intermediate viewing angles can expect,
through different lines of sight, can expect to sample contributions
from strata having different ionizations and velocities. In particular,
sight lines intersecting the star's visible pole and equator will tend
to run along streams having high and low velocities (and ionizations),
respectively. This may account for the variety of components along the
Na\,D and Balmer line profiles that are unlikely be caused by local 
ISM absorption, particularly for \va~ (see Fig.\,4a).  This might also 
well explain why the strongest broadenings are seen in metallic lines 
arising from intermediate and high excitations, such as the
Si\,II 6347, 6371\,\AA~ doublet and the infrared O\,I triplet.  
In summary, the wind/disk systems are difficult to reconcile with 
simple pictures of either: 
(1) line forming temperatures decreasing radially outward, or 
(2) an opening disk, as computed by Kraus \& Lamers (2003), with 
ionization contours streaming radially outward. The evidence supports 
a picture with components of both these geometries. However, in our
opinion the second picture probably is most consistent with the 
correlation of broadening of single absorption components with 
excitation potential.

  In addition to the complex geometrical and kinematical description,
the spatial distribution of the disks of \va~ and \vb~ offers
surprises with respect to the classical Be and the sgB[e] groups.
Consider first that forbidden emission lines are a defining attribute of 
B[e] and especially sgB[e] stars, and yet they are unobserved in these 
stars. This suggests, despite their large column density of some 10$^{24}$ 
cm$^{-2}$, and a likely typical density of 10$^{11-12}$ cm$^{-3}$,\footnote{
This inference comes from a modeling of the slow decrement of high-level 
Balmer absorption lines. A high N$_e$ is needed to permit $\tau_{line}$ 
to attain $\approx$ 1 and thus be visible.} 
that the disk does not extend to the lower densities ($\le$10$^{8-9}$ 
cm$^{-3}$), that favor forbidden line formation.  Consider further the 
relatively modest infrared excess for these stars of 
$\sim$1.7 magnitudes. This is relatively modest by standards of B[e] stars,
although it still falls within the sgB[e] distribution for these excesses.

  In all, the disks of the bgBe's are as thick and extensive in radius,
enough be visible in H$_{\alpha}$ out to 10's of or even 100 stellar radii. 
However, the evidence from the absence of [Fe] emission and only moderate 
IR emission suggests that the disk extents are limited and perhaps suddenly
truncated, such they do not show these expected additional traits.
One wonders what mechanism could truncate an otherwise continuing
gentle fall off in disk density with distance. As speculation, one 
possibility is a truncation occurs at the Lagrangian point in a widely
spaced binary system, particularly if the putative companion has a low mass. 
In this case, assuming a disk of extent 100R$_*$ (and thus an orbital radius 
of $\approx$5\,A.U.), a circular orbit viewed even edge-on would imply a 
radial velocity semi-amplitude of only a few km\,s$^{-1}$. Thus, a small RV
variation in a hypothetically wide spaced system could well
be undetectable for a small phase difference of $\le$0.2 cycles.

An even more puzzling attribute of these disks is the modulation
of starlight as a large condensation occulting the star once per
orbit. This hypothesis presents itself largely through the exclusion
of eclipsing binarity, which is untenable for several reasons, the 
improbability of observing 2 out of 2 widely spaced systems in eclipse
being among them.

\section{Conclusions}

In this paper we have investigated two photometrically variable SMC 
blue stars, viz.\,\va~ and \vb.~
These stars show rather exotic properties, such as CS absorption/emission 
lines, continuum reddening, an almost complete absence of photospheric lines, 
the presence of blue discrete absorption components (BACs) mostly in metallic 
lines, {\it and}  the absence of forbidden emission lines. We modeled FUV and 
high resolution optical lines with {\sc SYNSPEC} and {\sc circus} spectral 
synthesis codes finding that the spectra can be interpreted in terms  
of an optically thick slowly expanding and temperature-stratified wind/disk
star surrounding both stars. 
We interpreted discrete blue absorption components (``BACs") 
as evidence of a velocity discontinuity
in the outward flow of matter from the star, or more precisely as evidence
for weak shocks formed at the interface between a freely accelerating zone
close to the star and a very slowly expanding, flattened disk just beyond it.
We take this outer disk to be a relatively dense quasi-Keplerian structure
that is the source of H$\alpha$ emission and the rich metallic line spectrum.
We note a key feature, without being able to resolve why, and probably 
integral to their discovery, that the disks may not be axisymmetric. 
Nonaxisymmetric density bulges are reminiscent of the periodic migration
of one armed density waves, which are excited in the Be disks of many
classical Be stars (Okazaki 1991). The appearances of these disk features
manifest themselves
as periodic oscillations of the ratio $V/R$ in H$_{\alpha}$ emission
features in the Be stars. In light of the reversal of this ratio
between 2001 and 2007 for one of our program objects, one might
suspect that the ratios oscillate over time.
It is probably premature yet to suggest that this or other
alternative disk instability is responsible for the almost periodic
light curves until one can associate the light curve variations
more closely with disk properties.
Nonetheless, it is clear that this association could be
accomplished by monitoring the $V/R$ ratio in H$_{\alpha}$ in
these stars for the predicted variations over the next several
years.

   Although these objects are similar in some respects to sgB[e] stars, 
the program objects are unique kinds of Be stars by being less luminous 
than them and by displaying no forbidden emission lines in their spectra. 
Additionally, they have disks that are optically thick to many lines. 
They are also likely to have inhomogeneities 
which cause nearly periodic long variations in optical brightness.  
We speculate that these two objects are prototypes of a new type of 
variable - we have named them bgBe stars - with luminosities midway between
sgB[e] and Be stars. Although the bgBe stars are probably not close binary
systems, their membership in wide binaries with periods of a few years
or longer, cannot be ruled out. Wide binarity may even be favored to 
explain the absence of forbidden lines.

  To shed light on these new bgBe stars, optical spectroscopy is needed 
on other candidate objects discovered in our surveys of the Clouds. 
On the basis of extant lower quality spectra, 
we find at least three other stars in the LMC with properties similar 
to our two prototype objects. These are OGLE05155332-6925581, 
OGLE05141821-6912350, and OGLE00552027-7237101.  These spectra display 
moderate to strong H$\alpha$ emission profiles, high level Balmer 
absorption lines and variable (possibly regular) light curves. 
Although the resolution of these moderate-dispersion spectra is 
insufficient to see faint emissions, none of the spectra show
strong forbidden emission lines of Fe\,II. The spectrum of the first
of these is particularly interesting in that it shows BACs, and 
yet is a Doubly Periodic Variable (Mennickent et al. 2003, 
Mennickent et al. 2008). Because this object does not have a noticeable 
IR excess, its membership in a close binary could be important.  
Follow up observations of these and similar objects are ongoing and 
will help resolve such questions as the general prevalence of the bgBe
class, the dependence of their properties on metallicity and, for those 
members with thinner disks, their rotational velocities. Similarly,
the observation of UV resonance lines of these objects, so far lacking
in these objects, will elucidate the geometrical structure of their 
winds and perhaps allow the discovery of a high-velocity component.

\section{Acknowledgments}
REM acknowledges financial support by Fondecyt grant 1070705. 
We acknowledge G. Pietrzy\'nski for his help with the OGLE database. 
Our FUSE work was supported by NASA Grant NNX07AC75G to the
Catholic University of America. \\

\bsp 
\label{lastpage}
\end{document}